\documentclass[prc,aps,float,twocolumn,showpacs,nofootinbib,twoside]{revtex4}

\usepackage{amsmath}
\usepackage{amssymb}
\usepackage{amsthm}
\usepackage{amsfonts}
\usepackage{graphicx}
\usepackage[usenames,dvipsnames]{color}
\usepackage{afterpage}
\usepackage{mathrsfs}
\usepackage{graphicx,latexsym,epsfig}
\usepackage{mathrsfs}
\usepackage{color}
\usepackage{bm}% bold math
\usepackage[colorlinks,linkcolor=red,anchorcolor=blue,citecolor=blue]{hyperref}
\newcommand{\beq}{\vspace{0.5em}\begin{equation}}
\newcommand{\eeq}{\end{equation}\vspace{0.5em}}
\newcommand{\beqn}{\vspace{0.5em}\begin{eqnarray}}
\newcommand{\eeqn}{\end{eqnarray}\par\vspace{0.5em}\noindent}

\newcommand{\bsub}{\begin{subequations}}
\newcommand{\esub}{\end{subequations}}

\newcommand{\br}{{\mathbf{r}}}

\begin{document}

% \preprint{preprint}

\title{Global performance of multireference density functional theory for low-lying states in $sd$-shell nuclei}

\author{Xian-Ye Wu$^{1,3}$}
\author{Xian-Rong Zhou$^{1,2}$}\email[]{Corresponding author: xrzhou@phy.ecnu.edu.cn}

\affiliation{$^1$Department of Physics and Institute of Theoretical Physics and Astrophysics, Xiamen University, Xiamen 361005, China}
\affiliation{$^2$Department of Physics, East China Normal University, Shanghai, 200241, China}
\affiliation{$^3$School of Physical Science and Technology, Southwest University, Chongqing 400715, China}

%%%%%%%%%%%%%%%%%%%%%%%%%%%%%%%%%%%%%%%%%%%%%%%%%%%%%%%%%%%%%%%%

%\title{Global performance of multireference density functional theory for low-lying states in $sd$-shell nuclei}
%
%\author{X. Y. Wu}
%\affiliation{Department of Physics and Institute of Theoretical Physics and Astrophysics, Xiamen University, Xiamen 361005, China}
%\affiliation{School of Physical Science and Technology, Southwest University, Chongqing 400715, China}
%
%\author{X. R. Zhou}\email[]{Corresponding author: xrzhou@phy.ecnu.edu.cn}
%\affiliation{Department of Physics, East China Normal University, Shanghai, 200241, China}
%\affiliation{Department of Physics and Institute of Theoretical Physics and Astrophysics, Xiamen University, Xiamen 361005, China}
%\date{\today}

%%%%%%%%%%%%%%%%%%%%%%%%%%%%%%%%%%%%%%%%%%%%%%%%%%%%%%%%%%%%%%%%

%
 \begin{abstract}
 We present a comprehensive study of low-lying states in even-even Ne, Mg, Si, S, Ar isotopes with the multireference density functional theory (MR-DFT)  based on a relativistic point-coupling energy density functional (EDF). Beyond mean-field (BMF) effects are taken into account by configuration mixing of both particle-number and angular-momentum projected axially deformed states with generator coordinate method (GCM). Global performance of the MR-DFT for the properties of both ground state and of the first $2^+, 4^+$ states is examined, in comparison with previous studies based on nonrelativistic EDFs and available data. Our results indicate that an EDF parameterized at the BMF level is demanded to achieve a quantitative description.
 \end{abstract}

% 21.60.Jz:   Nuclear Density Functional Theory and extensions
% 21.10.Re:  Collective levels
% 27.50.+e:  59 ¡Ü A ¡Ü 89
%21.10.Ft
%21.10.Dr   Binding energies and masses
%21.10.Ft   Charge distribution
%21.10.-k   Properties of nuclei; nuclear energy levels (for properties of specific nuclei listed by mass ranges, see section 27)
%25.30.Bf   Elastic electron scattering
%25.30.Dh   Inelastic electron scattering to specific states
\maketitle
 \section{\label{introduction}Introduction}

 The combination of radioactive ion beans facilities and $\gamma$-ray detectors have allowed one to measure the low-lying excitation states of exotic nuclei far from $\beta$-stability. In the past decades, many interesting phenomena have been disclosed by studying the low-lying states of $sd$-shell nuclei, which cover three traditional neutron  magic numbers $N=8, 20, 28$~\cite{Mueller93,Tanihata95,Hansen95,Jonson04,Jensen04}.  Among them, the first $2^+$ and $4^+$ states are of particular interest since they provide rich information on the nuclear underlying shell structure and collective properties. The measured very low excitation energy of $2^+_1$ state and large transition strength $B(E2; 0_1^+\rightarrow2_1^+)$  in neutron-rich ${}^{32}$Mg  indicates the erosion of $N=20$ shell gap~\cite{Motobayashi95}. Similar phenomenon has been observed in ${}^{41,43}$P~\cite{Bastin07} and ${}^{40,42}$Si~\cite{Grevy04} which provide evidence for the collapse of $N = 28$ shell gap. These findings are consistent with other experimental measurements showing that the spherical $N=28$ shell gap is progressively reduced when more and more protons are removed from $^{48}$Ca~\cite{Sorlin93,Scheit96,Glasmacher97,Sohler02,Gade05,Grevy05,Gaudefroy06,Force10}.

 It is a challenge for nuclear models to describe the low-lying states of $sd$-shell nuclei, where the underlying shell structure changes rapidly when going from stable nuclei to dripline nuclei. In particular, clustering appears to be a common phenomenon in light nuclei~\cite{Arumugam05,Oertzen06,Beck11,Ebran14,Yao14b}. It complicates the naive pictures of  conventional shell model. In the pase decades, the model space of  shell model for $sd$-shell nuclei has been extended by including the $pf$ shell to take into account the effects of particle-hole excitations across the $N = 20$ shell gap, which are relevant to reproduce the low-lying states of neutron-rich nuclei in this mass region~\cite{Wildenthal80,Brown88,Utsuno99,Brown01,Otsuka01-3,Caurier05,Kaneko11}.

 In the framework of nuclear density functional theory (DFT) with a universal energy density functional (EDF) constructed phenomenologically based on the knowledge accumulated within modern selfconsistent mean-field models~\cite{Bender03}, the dynamical correlations related to the restoration of broken symmetries and to fluctuations of collective coordinates are very important for the low-lying states of $sd$-shell nuclei. These effects can be taken into account by implementation of symmetry-conserved generator coordinate method (GCM) into the DFT framework. This level of DFT is referred to as multireference (MR)-DFT, where the many-body energy takes the form of a functional of all mixed density matrices that are constructed from a chosen set of nonorthogonal configurations. In most of the studies, these configurations are generated by the selfconsistent mean-field calculations with constraints on nuclear multipole moments. For example, by adopting the configurations constrained to have axially symmetric shapes, the low-lying states of some selected isotopes in $sd$-shell have been studied~\cite{Rodriguez-Guzman00,Valor00,Bender03PRC,Niksic06}. In recent years, the MR-DFT approaches have been extended significantly by allowing for triaxially deformed configurations based on either non-relativistic~\cite{Bender08,Rodriguez2010} or relativistic EDFs~\cite{MRCDFT}, which provide a state-of-the-art microscopic calculation of nuclear low-lying states. Unfortunately, global application of the MR-DFT with triaxiality for nuclear low-lying states is still beyond the capabilities of current computers. Therefore, on one hand, a gaussian overlap approximation version of these approaches, i.e., a five-dimensional collective Hamiltonian (5DCH) method with the collective parameters determined from selfconsistent mean-field calculations is adopted for this purpose, either based on  the nonrelativistic finite-range Gogny force D1S~\cite{Bertsch07,Delaroche10} or the relativistic point-coupling PC-PK1 force~\cite{Lu15}. On the other hand, by assuming axial symmetry,  a global study of low-lying states for a large set of even-even nuclei has also been carried out  based on  the nonrelativistic Skyrme force SLy4~\cite{Bender05,Bender06a,Sabbey07} or the Gogny D1S and D1M forces~\cite{Rodr14}. In these global studies, however, little attention has been paid to the low-lying states of light nuclei.

 It is worth mentioning that the techniques of projection and GCM have also been implemented into the antisymmetrized molecular dynamics that uses a localized spherical~\cite{Kanada95} or triaxially deformed~\cite{Kimura04} Gaussian as the single-particle wave packet based on the Gogny force. This method has been applied to study many excited states in light nuclei and turns out to be very successful for describing clustering structures~\cite{Kanada12}. Besides, in recent years a shell-model-like multi-configuration approach has been developed based on the Gogny D1S force by mixing a set of symmetry-conserved orthogonal multiparticle-multihole configurations~\cite{Pillet08}. This method was examined very  recently by applying to study the low-lying states of $sd$-shell nuclei~\cite{Bloas14}. It has been shown that the standard deviation from experimental data is about 0.5 MeV for two-nucleon separation energies and about 0.4 MeV for excitation energies. However, the binding energies and $E2$ transition strengths were poorly described.

 In this work, we carry out a comprehensive study of $sd$-shell nuclei with a multireference covariant DFT~\cite{Yao13,Yao14}, which is an extended version of the model that has been applied to the low-lying states of carbon~\cite{Yao11-2} and magnesium isotopes~\cite{Yao11} by implementing an  additional technique of particle number projection. For simplicity, the configurations of the present global study are restricted to have axial symmetry considering the fact that the triaxiality effect turns out to be marginal in the ground-state~\cite{Wang14} and the first $2^+, 4^+$ states in light nuclei~\cite{Yao11-2,Yao11}. To examine the global performance of MR-DFT approach, the results are discussed in comparison with previous similar studies based on nonrelativistic EDFs and available data. The purpose of this work is to address the following questions: (1) How are the nuclear  properties modified by the  beyond-mean-field (BMF) effect of projections and configuration mixing? (2) How good is the MR-DFT for the low-lying states of light nuclei?

 The paper is arranged as follows. In Sec.~\ref{Sec.II}, we present a brief introduction to the framework of the MR-CDFT. In Sec.~\ref{Sec.III}, the ground-state properties including binding energy, separate energy, density distribution, charge radii, and the excitation energies of $2^+_1, 4^+_1$ states, electric quadrupole transition strengths from $0^+_1$ state to $2^+_1$ state, spectroscopic quadrupole moments and neutron-proton decoupling factors of the $2^+_1$ state are discussed in comparison with available data.  A summary of our findings and an outlook are given in Sec.~\ref{Sec.IV}.

 \section{The multireference covariant density functional theory}%
 \label{Sec.II}
 \subsection{Relativistic mean-field calculation with point-coupling effective interaction}
 We start from a nonlinear point-coupling effective Lagrangian that determines the energy functional of a nuclear system in terms of local single-nucleon densities and currents
 \begin{eqnarray}
 {{E}}_{\rm RMF}
 &=& \int d{\bm r }~{\mathcal{E}_{\rm RMF}}(\bm{r})\nonumber \\
 &=&\int d\bm{r} \sum_k{~v_k^2
 ~{\bar{\psi}_k (\bm{r}) \left( -i\bm{\gamma}\bm{\nabla} + m\right )\psi_k(\bm{r})}}\nonumber \\
 &+& \int d{\bm r }~{\left(\frac{\alpha_S}{2}\rho_S^2+\frac{\beta_S}{3}\rho_S^3 +
 \frac{\gamma_S}{4}\rho_S^4+\frac{\delta_S}{2}\rho_S\triangle \rho_S \right.}\nonumber \\
 &+&  {\left.\frac{\alpha_V}{2}j_\mu j^\mu + \frac{\gamma_V}{4}(j_\mu j^\mu)^2 +
 \frac{\delta_V}{2}j_\mu\triangle j^\mu \right.} \nonumber \\
 &+& \left. \frac{\alpha_{TV}}{2}j^{\mu}_{TV}(j_{TV})_\mu+\frac{\delta_{TV}}{2}
 j^\mu_{TV}\triangle  (j_{TV})_{\mu}\right.\nonumber \\
 &+&\frac{\alpha_{TS}}{2}\rho_{TS}^2
 \left.+\frac{\delta_{TS}}{2}\rho_{TS}\triangle
 \rho_{TS} +\frac{e}{2}j^{\mu}_p A_\mu
 \right) ,
 \label{EDF}
 \end{eqnarray}
 where the coupling constants  $\alpha_i, \beta_i, \gamma_i, \delta_i$ are determined in the optimization of the EDF for the properties of several finite nuclei and nuclear matter~\cite{Burvenich02,Zhao2010,Meng2013}. $A_\mu$ is the four-component electromagnetic field, and the densities $\rho_i$ and currents  $j^\mu_i$ are bilinear
 combinations of Dirac spinor field of nucleon, namely $\bar\psi\Gamma_i\psi$
 with $i=S, V, TV$ representing the symmetry of the coupling.
 The subscript $S$ stands for isoscalar-scalar ($\Gamma_S = 1$), $V$
 for isoscalar-vector  ($\Gamma_V = \gamma^\mu$), and $TV$ for
 isovector-vector ($\Gamma_{TV} =\gamma^\mu t_3$) type of coupling
 characterized by their transformation properties in isospin and
 in space-time. The densities and currents
 \begin{subequations}
 \begin{align}
 \label{dens_1}
 \rho_{S}({\bm r}) &=\sum_k v_k^2 ~\bar{\psi}_{k}({\bm r})
              \psi _{k}({\bm r})~,  \\
 \label{dens_2}
 \rho_{TS}({\bm r}) &=\sum_k v_k^2 ~
       \bar{\psi}_{k}({\bm r})\tau_3\psi _{k}^{{}}({\bm r})~,  \\
 \label{dens_3}
 j^{\mu}({\bm r}) &=\sum_k v_k^2 ~\bar{\psi}_{k}({\bm r})
         \gamma^\mu\psi _{k}^{{}}({\bm r})~,  \\
 \label{dens_4}
 j^{\mu}_{TV}({\bm r}) &=\sum_k v_k^2 ~\bar{\psi}_{k}({\bm r})
      \gamma^\mu \tau_3 \psi _{k}^{{}}({\bm r})~,
 \end{align}
 \end{subequations}
 are calculated in the {\it no-sea} approximation, i.e., the summation in
 Eqs.~(\ref{dens_1}) - (\ref{dens_4}) runs over all occupied states in
 the Fermi sea. $v_k^2$ denotes the occupation probability of the $k$-th single-nucleon state.

 Because of charge conservation, only the $3$rd component of the isovector densities and currents contributes to the
 nucleon selfenergies. In this work we only consider even-even nuclei, i.e., time-reversal invariance is assumed, which implies that the spatial components of the single-nucleon currents vanish in the mean-field states \cite{Meng1998,Long2004,Meng2006}. The single-nucleon wave functions are obtained as selfconsistent solutions of the Dirac equation
 \begin{eqnarray}
 \label{Dirac:N}
 \left[ \bm{\alpha}\cdot\bm{p}+V_0({\bm r}) +\beta \big(m+S({\bm r})\big)\right]\psi_{k}({\bm r})=\epsilon_k\psi_{k}({\bm r})\, ,
 \end{eqnarray}
 where the scalar and vector potentials
%%%%%%%%%%%%%%%%%%%%%%%%%%%%%%%%%%%%%%%%%%%%%%%%%%%
 \begin{equation}
 S({\bm r}) = \Sigma_S({\bm r}) + \tau_3\Sigma_{TS}({\bm r})\;,
 \label{scapot}
 \end{equation}
%%%%%%%%%%%%%%%%%%%%%%%%%%%%%%%%%%%%%%%%%%%%%%%%%%%
 \begin{equation}
 V^{\mu}({\bm r})  = \Sigma^{\mu}({\bm r}) + \tau_3\Sigma^{\mu}_{TV}({\bm r})\;,
 \label{vecpot}
 \end{equation}
 contain the nucleon isoscalar-scalar, isovector-scalar,
 isoscalar-vector and isovector-vector self-energies
 defined, respectively, by the following relations
 \begin{subequations}\begin{eqnarray}
    \label{selfS}
    \Sigma_S & = & \alpha_S \rho_S + \beta_S \rho_S^2 +
    \gamma_S\rho_S^3+ \delta_S \triangle \rho_S \; ,\\
    \label{selfTS}
    \Sigma_{TS} & = & \alpha_{TS} \rho_{TS}
    +\delta_{TS} \triangle \rho_{TS}\; , \\
    \label{selfV}
    \Sigma^{\mu} & = & \alpha_V j^{\mu}
   +\gamma_V (j_\nu j^\nu)j^\mu + \delta_V \triangle j^\mu
    -eA^\mu\frac{1-\tau_3}{2} \; ,\nonumber \\ \\
    \label{selfTV}
    \Sigma^{\mu}_{TV} & = & \alpha_{TV} j^{\mu}_{TV}
       + \delta_{TV} \triangle j^{\mu}_{TV}\, .
 \end{eqnarray}\end{subequations}

  To generate the mean-field wave functions $\vert \Phi(\beta)\rangle$ with different deformation parameters $\beta, \gamma$, a quadratic constraint on the mass quadrupole moments is added in the variation of the energy function
  \begin{equation}
  E_{\rm RMF} + \sum_{\mu=0, 2} C_{2\mu}(\langle \hat{Q}_{2\mu}\rangle - q_{2\mu})^2\, ,
  \label{constraint}
  \end{equation}
  where $C_{2\mu}$ is a stiffness parameter and $\langle \hat{Q}_{2\mu}\rangle$ denotes the expectation value of the mass quadrupole moment operator
  \begin{equation}
   \hat{Q}_{20}  =  \sqrt{\dfrac{5}{16\pi}} (2z^2 - x^2 - y^2)\, ,~~
    \hat{Q}_{22}  = \sqrt{\dfrac{15}{32\pi}} (x^2 - y^2)\, .
  \end{equation}
  In Eq. (\ref{constraint}), $q_{2\mu}$ is the quadrupole moment of mean-field state to be obtained. Since we are restricted to axially deformed configurations, the $q_{22}$ is set to zero, corresponding to the deformation parameter $\gamma=180^\circ$ and $0^\circ$. The deformation parameter $\beta$ is related to the expectation values of the mass quadrupole moment operator by $\beta= \dfrac{4\pi}{3AR^2} \langle \hat Q_{20}\rangle$  with $R=1.2A^{1/3}$ (fm).

  \subsection{Beyond mean-field calculation with symmetry-restored generator coordinate method}
  The wave function of nuclear low-lying states is given by configuration mixing of angular-momentum and particle-number projected axial relativistic mean-field wave functions
  \beqn%
  \label{wavefun}
  \vert \Psi^{JNZ}_\alpha\rangle
  =\sum_\beta f^{J}_\alpha(\beta)\hat P^J_{MK=0} \hat P^N\hat P^Z\vert \Phi(\beta)\rangle  \, ,
  \eeqn
  where $\alpha$ labels different collective states for a given angular
  momentum $J$.  The $\hat P^N$, $\hat P^Z$ and $\hat P^J_{MK}$  are projection operators onto neutron number $N$, proton number $Z$ and angular momentum $J$, respectively.
  The $\vert\Phi(\beta)\rangle$s are a set of Slater determinants of quasiparticle wavefunctions from the previous deformation constrained RMF+BCS calculation. The weight functions
  $f^{J}_\alpha(\beta)$ are determined by minimizing the energy of the collective state with respect to the weight function. This leads to the Hill-Wheeler-Griffin
  equation~\cite{Hill53,Griffin57,Ring80}
  \beqn%
 \label{HWGE}
 \sum_{\beta^\prime} \left[ \mathcal{H}^{J}(\beta,\beta^\prime)
          - E_\alpha^{J} \, \mathcal{N}^{J}(\beta,\beta^\prime)
           \right] f_\alpha^{J}(\beta^\prime) = 0 \, .
  \eeqn%
  where $\mathcal{N}^{J}(\beta,\beta^\prime)=\langle\Phi(\beta)\vert
  \hat P^J_{00}\hat P^N\hat P^Z\vert \Phi(\beta)\rangle$ and
  $\mathcal{H}^{J}(\beta,\beta^\prime)=\langle\Phi(\beta)\vert\hat H\hat
  P^J_{00}\hat P^N\hat P^Z\vert \Phi(\beta)\rangle$ are the energy kernel
  and the norm kernel, respectively. In our calculation, the prescription of
  mixed densities is used to calculate the energy kernel~\cite{Yao10}. As the
  projected mean-field states do not form an orthogonal basis and the expansion
  coefficients $f^{J}_\alpha(\beta)$ in Eq. (\ref{wavefun}) are not orthogonal,
  a set of orthonormal collective wave functions $g_\alpha^J(\beta)$ related with
  weight function is usually constructed~\cite{Ring80}
   \beqn%
   \label{gfunc}
   g_\alpha^J(\beta)=\sum_{\beta^\prime}\left[\mathcal{N}^{J}
   \right]^{1/2}(\beta,\beta^\prime) f_\alpha^{J}(\beta^\prime)\, ,
   \eeqn%
  which provide the information of dominated configurations in the collective
  states $\vert \Psi^{JNZ}_\alpha\rangle$. The solution of the Hill-Wheeler-Griffin
  equation provides the weight functions $f^{J}_\alpha(\beta)$
  and the energy spectrum, as well as other information needed for calculating
  the electric multipole transition strengths~\cite{Ring80}.
  More details to the MR-CDFT have been introduced in Ref.~\cite{MRCDFT}.

 \section{Results and discussions}
 \label{Sec.III}

 In the relativistic mean-field calculation, parity, time-reversal invariance and axial symmetry are imposed. The Dirac equation (\ref{Dirac:N}) is solved by expanding the nucleon wave function in terms of eigenfunctions of a three-dimensional harmonic oscillator in Cartesian coordinate with 10 major shells~\cite{Yao06}. The relativistic point-coupling forces of both PC-F1~\cite{Burvenich02} and PC-PK1~\cite{Zhao2010} are employed for comparison. Pairing correlations between nucleons are treated with the BCS approximation using a density-independent $\delta$ force with a smooth cutoff factor~\cite{Krieger90}. In the beyond-mean-field calculation, the Gauss-Legendre quadrature is used for the integrals over the Euler angle $\theta$ in the calculation of the projected kernels. The number of mesh points in the interval $\theta\in[0,\pi]$ is chosen as $N_\theta=10$. The number of gauge angles in the Fomenko's expansion~\cite{Fomenko70} for the particle number projection is $N_\phi=9$. The Pfaffian method~\cite{Robledo09} has been implemented to calculate the norm overlap. Like other multireference density functional methods, the problems of self-interactions and self-pairing~\cite{Bender09} might exist in our method. However, we have not seen any evidence that these problems are very serious to be taken into account. We have checked our calculation against the number of mesh points in the Gauge angle and obtained a good plateau condition. All the results are obtained based on this plateau condition.

 \subsection{An illustrative calculation for $^{44}$S}

  \begin{figure}[]
  \centering
  \includegraphics[width=6.5cm]{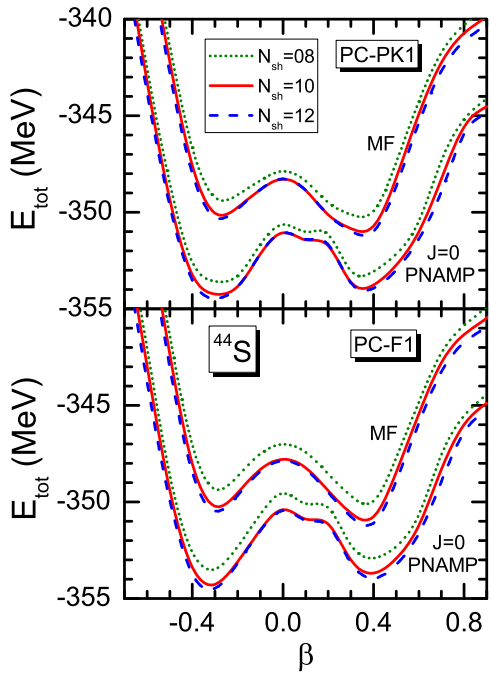}
  \caption{(Color online) The energy of mean-field states in  $^{44}$S and that of projected states with projection onto particle number and angular momentum ($J=0$) as functions of axial deformation parameter $\beta$. The calculations are performed by expanding the Dirac spinors of nucleons in the harmonic oscillator basis with different number of major shells.}
  \label{fig:PECsS44}
  \end{figure}

   We first examine this method for the shape-coexistence nucleus $^{44}$S, which has attracted a lot of attention in recent years concerning the erosion of $N=28$ shell gap~\cite{Force10,Li11,Tomas11,Santiago-Gonzalez11,Caceres12,Utsuno2015}. In the framework of DFT, an oblate deformed and $\gamma$-soft $0^+$ as well as a prolate deformed $0^+_2$ states are obtained in the 5DCH calculation based on the relativistic DD-PC1 force~\cite{Li11} and in the MR-DFT calculation based on the nonrelativistic Gogny force~\cite{Tomas11}. It is interesting to know how good is the description of the present MR-DFT calculation for the low-lying states of $^{44}$S with only axially deformed configurations based on the relativistic EDFs.

\begin{figure}[]
  \centering
  \includegraphics[width=8.5cm]{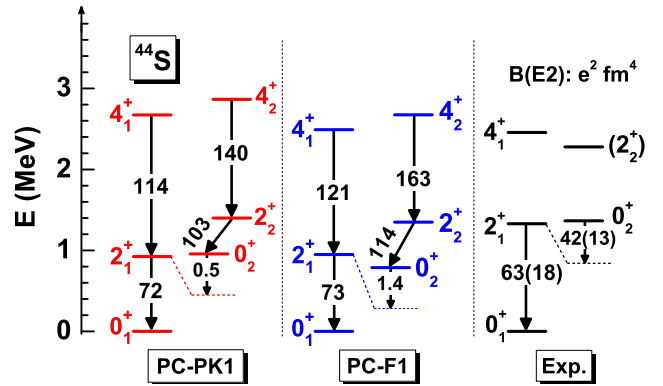}
  \caption{(Color online) Low-lying spectra of $^{44}$S calculated with the MR-CDFT using the PC-PK1 and PC-F1 forces, in comparison with available data~\cite{Force10}. The numbers on the arrow are the $B(E2)$ values (in units of $e^2$ fm$^4$).}
  \label{fig:specS44}
  \end{figure}

   Figure~\ref{fig:PECsS44} displays the mean-field and projected ($N, Z, J=0$) energies of the axially deformed mean-field configurations in $^{44}$S as a function of the intrinsic deformation $\beta$ calculated using the PC-PK1 and PC-F1 force with different number of harmonic oscillator major shells. It is shown that the 10 major shells are sufficient to provide convergent results on energy. Fig.~\ref{fig:specS44} shows the low-lying spectra of $^{44}$S calculated with both PC-PK1 and PC-F1 forces, in comparison to available data. The calculated excitation energy of the first $2^+$ excited state is $\sim30\%$ lower than experiment. However, the $4^+_1$ excitation energy shows much closer to the datum. For the transition strengths, the experiment $B(E2; 2^+_1 \to 0^+_1)$ is reproduced. One can also notice that the $E2$ transition strength between $0^+_2$ and $2^+_1$ states is significantly underestimated by both forces. The previous study of magnesium isotopes with triaxiality~\cite{Yao11} shows that the excitation energies of $2^+_1$ and $4^+_1$ states are not significantly changed by the triaxiality. However, the triaxiality effect may show up at some nuclei, like $^{44}$S. To reduce computational burden, we simply neglect the triaxiality effect in the present systematic calculation.

  \subsection{Ground-state properties}

  \subsubsection{Binding energy and shell gap}

  \begin{figure}[]
  \centering
  \includegraphics[width=6.0cm]{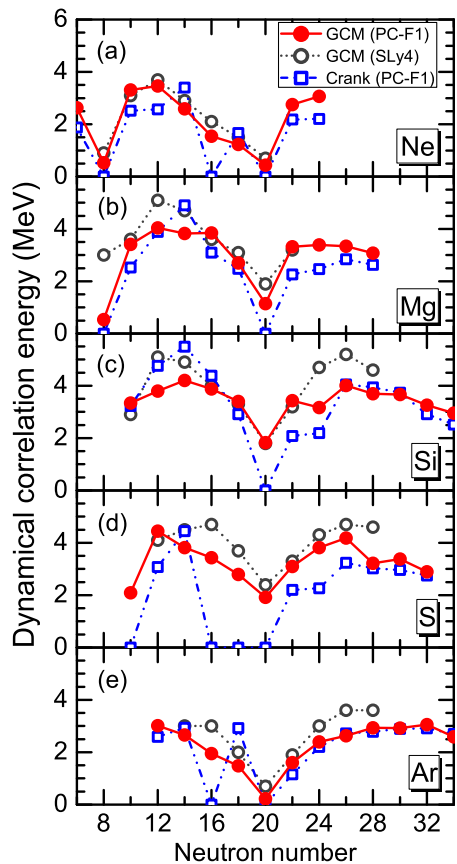}
  \caption{(Color online) Dynamical correlation energies of $sd$-shell nuclei from the configuration-mixing GCM calculation using the PC-F1 force, in comparison with the similar calculation using the Skyrme SLy4 force~\cite{Bend06}. The results by the cranking prescription based on the mean-field ground-state solution by the PC-F1 force are also given for comparison. }
  \label{fig:DCE}
  \end{figure}

  The energy of ground state ($0^+_1$) can be divided into two parts
   \beqn
   \label{Ecorr}
   E(0^+_1) = E_{\rm NZ}(\beta_m)+E_{\rm Dyn}\, ,
   \eeqn%
  where $E_{\rm NZ}(\beta_m)$ denotes the energy of the global minimum (with intrinsic deformation $\beta_m$) on the particle-number projected energy curve and the $E_{\rm Dyn}$ is dynamical correlation energy gained from symmetry restoration and shape mixing. The dominated part of  $E_{\rm Dyn}$ is from angular momentum projection. Therefore, for simplicity, this dynamical correlation energy has usually been taken into account phenomenologically with the cranking prescription~\cite{Gori05,Cham08,Gori09a,Zhang14}
  \begin{equation}
  \label{dyn-cranking}
  E^{\rm Crank}_{\rm Dyn}=E_{\rm rot}\{b{\rm tanh}(c|\beta_m|)+d|\beta_m|{\rm e}^{-l(|\beta_m|-\beta^{0})^{2}}\}\, ,
  \end{equation}
 where $E_{\rm rot}$ is so-called rotational correction energy
  \begin{equation}
  \label{rot}
  E_{\rm rot}=\frac{\hbar^{2}}{2{\cal I}}\langle\hat{J}^{2}\rangle\, ,
  \end{equation}
  with the moment of inertia ${\cal I}$ calculated by the Inglis-Belyaev formula and with $\hat{J}$ being the angular momentum operator.  The parameters $b, c, d, l, \beta^0$ are optimized to the nuclear masses in Ref.~\cite{Cham08}, where the values of 0.80, 10, 2.6, 10, 0.10 are obtained for these parameters, respectively. This prescription turns out to be very successful to improve nuclear mass models. A better treatment of the dynamical correlation energy is to carry out the calculation with exact quantum-number projections in the framework of GCM, which has been done based on the Skyrme force~\cite{Bend05,Bend06} or based on the Gogny forces~\cite{Rodr14}. Both calculations are restricted to axially deformed configurations. It is worth mentioning that by solving the 5DCH with triaxiality,  the dynamical correlation energy is evaluated by using either the Gogny D1M~\cite{Gori09b} and D1S~\cite {Delaroche10} forces or the relativistic PC-PK1 force~\cite{Lu15}.

  To examine the validity of the cranking prescription in Eq. (\ref{dyn-cranking}) against the exact projection plus GCM calculation, we make a comparison of the dynamical correlation energies calculated in these two ways based on the same energy functional PC-F1. The results are plotted in  Fig.~\ref{fig:DCE} as a function of neutron number. According to the cranking prescription~\cite{Zhang14}, the dynamical correlation energy is zero if the mean-field solution is spherical, and is large for well-deformed mean-field solution. This is not the case in the exact projection plus GCM calculation. Even for spherical vibrator nuclei, there are still some amount of dynamical correlation energies. This nonzero correction energy is mainly originated from shape mixing. Fig.~\ref{fig:DCE} shows that the largest discrepancy in the dynamical correlation energies between the cranking prescription and the GCM calculations is found in spherical vibrational nucleus $^{32}$S with energy difference 3.5 MeV. Since the dynamical correlation energy is sensitive to nuclear deformation and the underlying shell structure, it provides an indirect way to unveil the shell structure evolution towards dripline, such as the erosion of $N=28$ shell gap in neutron-rich nuclei. To examine the force-dependence of the dynamical correlation energy, we also compare the values from the GCM calculation using the SLy4~\cite{Bend06} in Fig.~\ref{fig:DCE}. The results given by these two different forces are similar for most cases except for $^{20}$Mg.

  \begin{figure}[]
  \centering
  \includegraphics[width=8cm]{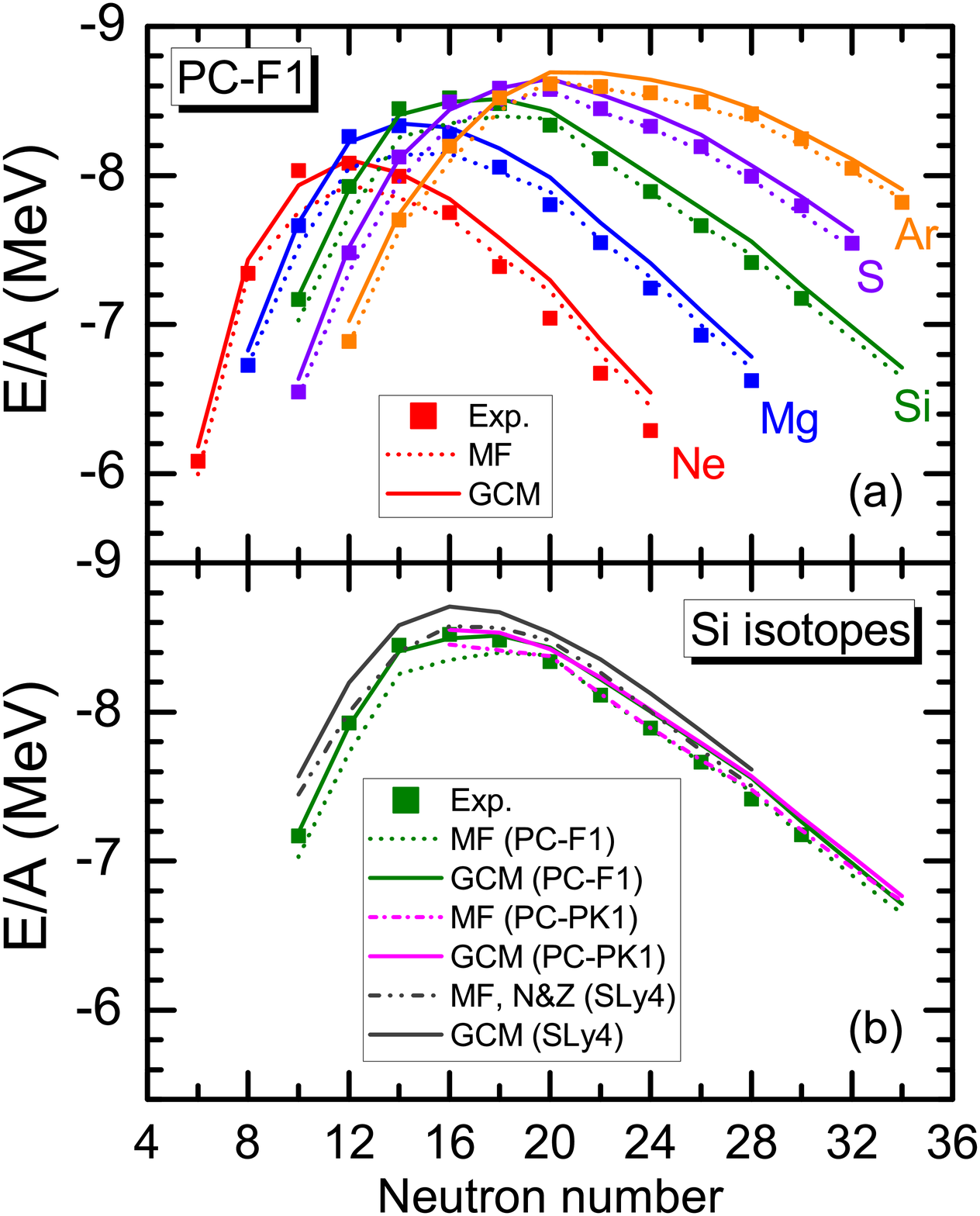}
  \caption{(Color online) (a) Energy per nucleon for Ne, Mg, Si, S, and Ar isotopes
  in $sd$-shell region from both mean-field and configuration-mixing GCM calculations using the PC-F1 force, in comparison with data. (b) Energy per nucleon for silicon isotopes  calculated using the relativistic PC-F1 and PC-PK1 forces in comparison with those by the nonrelativistic SLy4 force from Ref.~\cite{Bend06} and available data from Ref.~\cite{Audi03}. The MF results of nonrelativistic SLy4 include the correlation energy gained from the particle number projection.}
  \label{fig:sdBE}
  \end{figure}

  Figure~\ref{fig:sdBE} shows the energy per nucleon by both mean-field and configuration-mixing GCM calculations in comparison with available data~\cite{Audi03}. It is shown that mean-field calculation underestimate the average binding energy for neutron-deficient nuclei. The inclusion of the dynamic correlation energies thus improves the description of the energies for these nuclei. In the mean time, however, one reads an overestimation of the binding energies for neutron-rich nuclei. It leads to increase the rms deviation of binding energies of 56 $sd$-shell nuclei from 2.94 MeV to 3.61 MeV on beyond-mean-field calculations. Similar phenomenon is also observed in the MR-DFT calculation using the PC-PK1 and SLy4 forces. This problem can be traced back to the way that the energy density functionals were optimized at the mean-field level. A better description of the masses of the neutron-rich nuclei asks for a new energy functional parameterized at the beyond-mean-field level, as it was done for the Gogny D1M force~\cite{Gori09b}.

  \begin{figure}[]
  \centering
  \includegraphics[clip=,width=7.5cm]{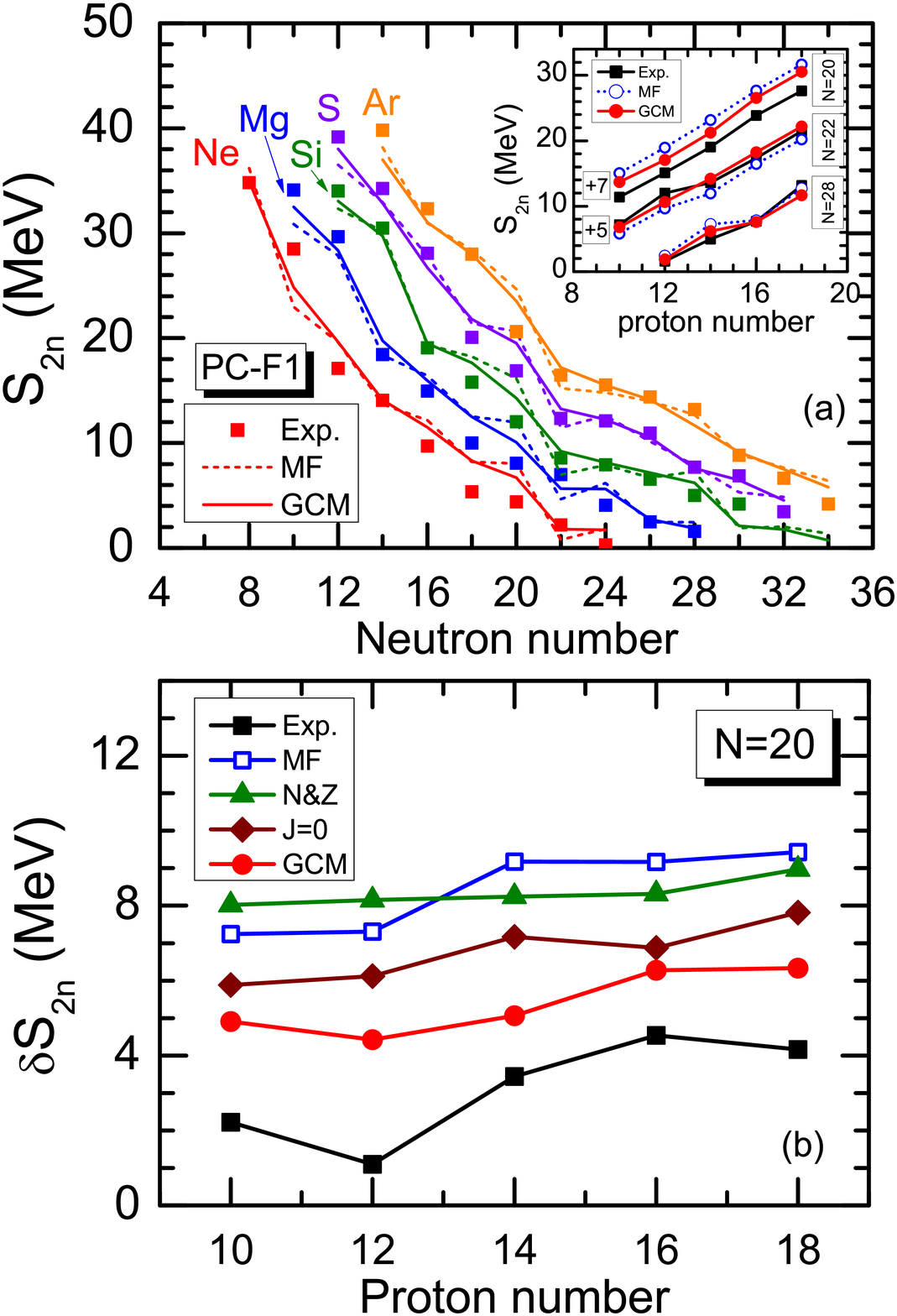}
   \caption{(Color online) (a) Two-neutron separation energy as a function of neutron number.  The inset panel shows
   that of $N=20$, 22 and 28 isotonic chains. (b) Neutron $N=20$ shell gap  by the mean-field (MF), PNP ($N\&Z$), PN1DAMP ($J=0$) and full GCM calculations using the  PC-F1 force as a function of proton number.  Data are taken from Ref.~\cite{Audi03}. }
  \label{fig:sdS2n}
  \end{figure}

  The nucleon separation energy is usually better described than the binding energy after taking into account dynamical correlation energy~\cite{Wang14,Lu15}. This phenomenon is also shown in Fig.~\ref{fig:sdS2n}, where the two-neutron separation energies $S_{2n}=E(Z, N-2 )- E(Z, N)$ from both the mean-field and GCM calculation using the PC-F1 force are plotted as a function of neutron number, in comparison with data~\cite{Audi03}. The inset of Fig.~\ref{fig:sdS2n}(a) shows that the neutron $N=20, 22$ and $N=28$  shell gaps are better described after taking into account the dynamical correlation energies. The information of neutron shell gap can be learnt in a quantitative way by defining a differential of two-nucleon separation energy $\delta S_{2\mathrm{n}}(Z, N)
  = E(Z, N-2) + E(Z, N+2) - 2E(Z, N)$, where $E(Z, N)$ is total energy of the nucleus with proton number $Z$ and neutron number $N$, respectively. Fig.~\ref{fig:sdS2n}(b) displays in detail how the $N=20$ shell gap is better described step by step. For this purpose, the results from the mean-field, particle number projection, particle number and angular momentum projection, as well as configuration mixing GCM calculations using the  PC-F1 force are compared with available data. It is shown that the $N=20$ shell gap is better described after taking into account more and more correlations. To know the accuracy of global description, we calculate the root-mean-square (rms) deviation $\delta\equiv\sqrt{\frac{1}{N}\sum_i(O^{cal}_i-O^{exp}_i)^2}$ for a set of 51 two-neutron separation energies, and obtain the $\delta$ value as $\sim2.07$ MeV and $\sim1.53$ MeV from the mean-field and configuration-mixing calculations, respectively.

  \subsubsection{Proton and charge distributions}

  With the weight function, it is straightforward to evaluate the the density distribution of the low-lying states $J^+_\alpha$ in laboratory frame with the information of shape mixing~\cite{Yao12,Yao13,Yao15}
  \begin{eqnarray}
  \label{eqGCM}
  \rho^{J\alpha}(\br)
  & = &   \sum_{\beta\beta'}  f^{J}_{\alpha} (\beta')f^{J}_\alpha(\beta)
  \sum_{\lambda}  (-1)^{2\lambda}Y_{\lambda0}(\hat \br) \nonumber\\
  &&\times
  \langle  J0,\lambda 0\vert J 0\rangle  \sum_{K}(-1)^{K}\langle JK,\lambda -K\vert J0\rangle \nonumber \\
        &   & \times
  \int d\hat\br^\prime \rho^{JK0}_{\beta'\beta}(\br^\prime)Y^\ast_{\lambda K}(\hat \br^\prime) \, ,
  \end{eqnarray}
  where $\rho^{JK0}_{\beta'\beta}(\br)$ is defined as
  \begin{eqnarray}
  \label{rho_JKNZ}
  \rho^{JK0}_{\beta'\beta}(\br)
  &\equiv& \dfrac{2J+1}{2} \int^\pi_0 d\theta\sin(\theta) d^{J\ast}_{K0}(\theta)\nonumber\\
  &&\times
  \langle  \Phi(\beta^\prime) \vert \sum_{i}\delta(\br-\br_i) e^{i\theta\hat J_y}
  \hat P^{N}\hat P^{Z}  \vert \Phi(\beta)\rangle \, .\nonumber\\
  \end{eqnarray}
  the index $i$ in the summation runs over all the occupied single-particle states for neutrons or protons. $\hat\br\equiv(r, \hat \br)$ is the position at which the density is to be calculated. $\br_i$ is the position of the $i$-th nucleon.

  We note that the density by Eq. (\ref{eqGCM}) contains the information of many deformed mean-field states generated by the quadrupole deformation $\beta$. For the ground state $0^+_1$, the density is simplified as
  \begin{eqnarray}
  \rho^{01}(r) =  \sum_{\beta\beta'}  f^{0}_1 (\beta')f^{0}_1(\beta) \int d\hat\br\rho^{000}_{\beta'\beta}(\br)\, .
  \end{eqnarray}

  \begin{figure}[]
  \centering
  \includegraphics[width=8.5cm]{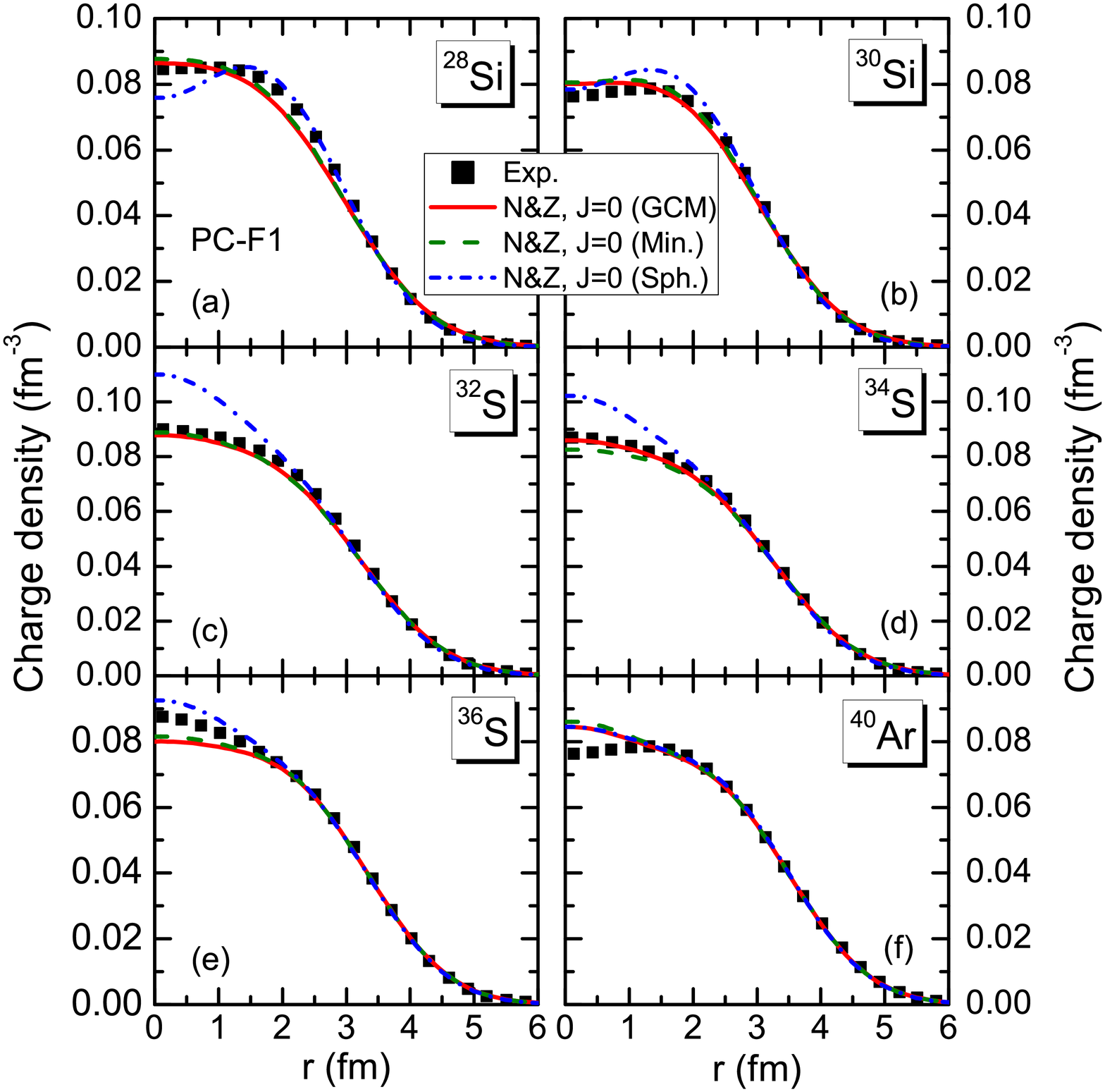}
  \caption{(Color online) Charge density distributions calculated
  with the spherical configuration, the configuration corresponding to the global
  minimum of $J=0$ energy curve (Min.), and the full configurations (GCM), respectively.
  The charge density units are in fm$^{-3}$. The experimental data from Ref.~\cite{Richter03}
  are given for comparison.}
  \label{fig:expdens}
  \end{figure}

\begin{figure}[]
  \centering
  \includegraphics[width=6.5cm]{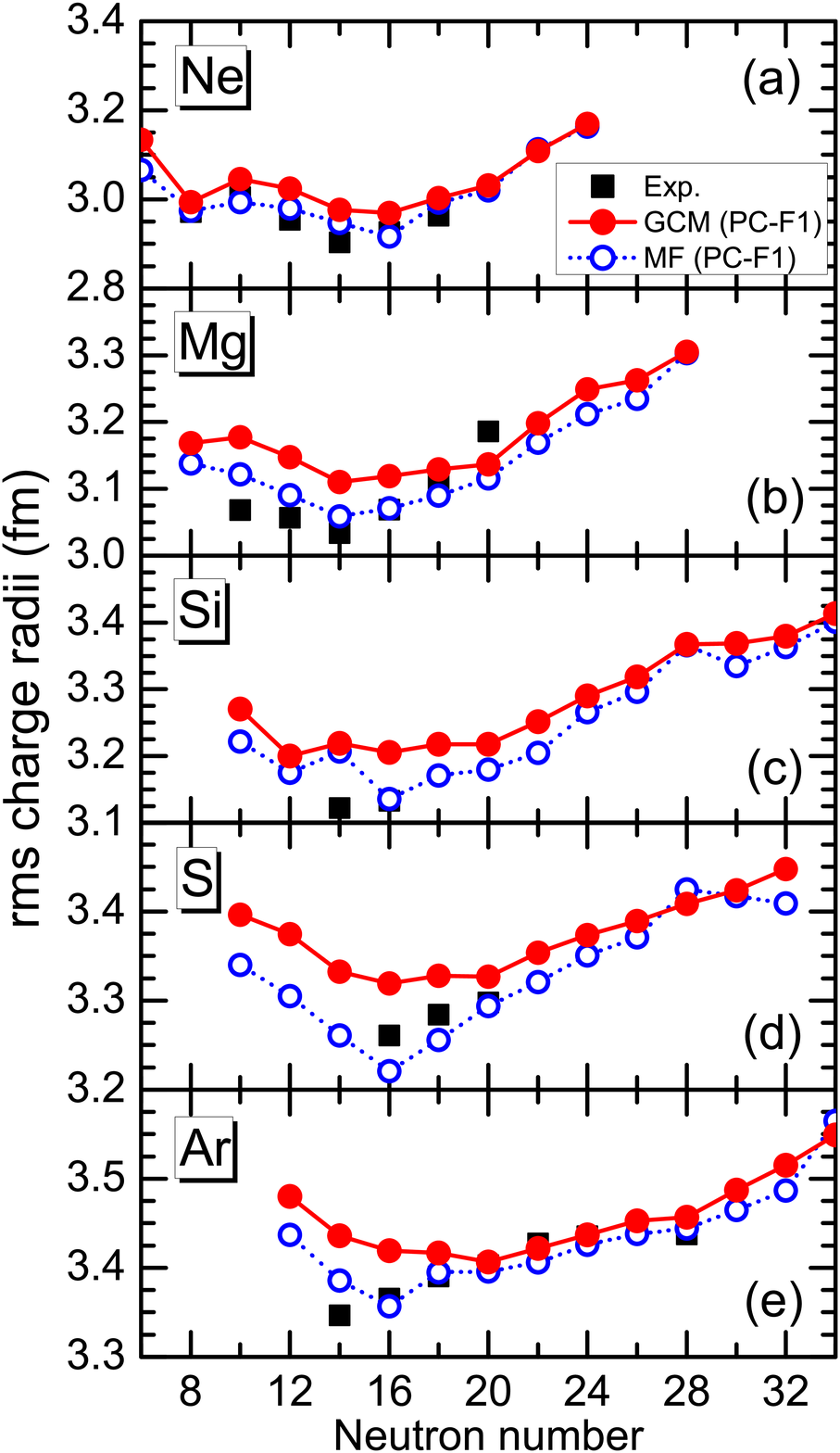}
  \caption{(Color online) Root-mean-square charge radii of $sd$-shell even-even nuclei as
  functions of neutron number. The results for Ne, Mg, Si, S, and Ar isotopes obtained
  by PN1DAMP+GCM (red line) are compared with those from the RMF+BCS mean-field (blue line)
  using the same relativistic energy density functional. The available data for Mg isotopes
  from Ref.~\cite{Yordanov12} and others from Ref.~\cite{Angeli13} are also given
  for comparison.}
  \label{fig:rmschar}
  \end{figure}

  Figure~\ref{fig:expdens} displays the charge densities in $^{28, 30}$Si, $^{32, 34, 36}$S and $^{40}$Ar as a function of radial coordinate $r$ from three types of calculations with different configurations based on the PC-F1 force, in comparison with  data. The charge density is calculated by convolution of the corresponding proton density with a Gaussian form factor
  \beqn
  \rho_{\rm ch}(r)
  &=&\frac{1}{a\sqrt{\pi}}\int dr^\prime r^\prime\rho_{p}(r^\prime)\nonumber\\
  &&\times\left[\frac{e^{-(r-r^\prime)^2/a^2}}{r}
  -\frac{e^{-(r+r^\prime)^2/a^2}}{r}\right],
  \label{rhocha}
  \eeqn
  where the parameter is chosen as $a=0.65$ fm.  The charge density of particle number conserved spherical state in $^{28}$Si shows a somewhat center depletion, in contrast to experimental data. This phenomenon is also observed in the Hartree-Fock calculation using the Skyrme Skm* force~\cite{Richter03}. After taking into account static and dynamic quadrupole deformation effects, the center depletion is totally washed out and becomes closer to the data in the interior, except for pushing the density in the middle (from $\sim$1.3 fm to $\sim$2.0 fm) to nuclear surface. It leads to an overestimation of rms charge density in $^{28}$Si by $\sim0.1$ fm (cf. Fig.~\ref{fig:rmschar}). A much better agreement with data is observed in the charge densities of $^{30}$Si and $^{32-36}$S. However, Fig.~\ref{fig:rmschar} shows that the rms charge radii of $^{32-36}$S are overestimated. The static and dynamic quadrupole deformation effects turn out to be negligible in the charge distribution and rms charge radius of $^{40}$Ar.

\begin{figure}[]
  \centering
  \includegraphics[clip=,width=7.5cm]{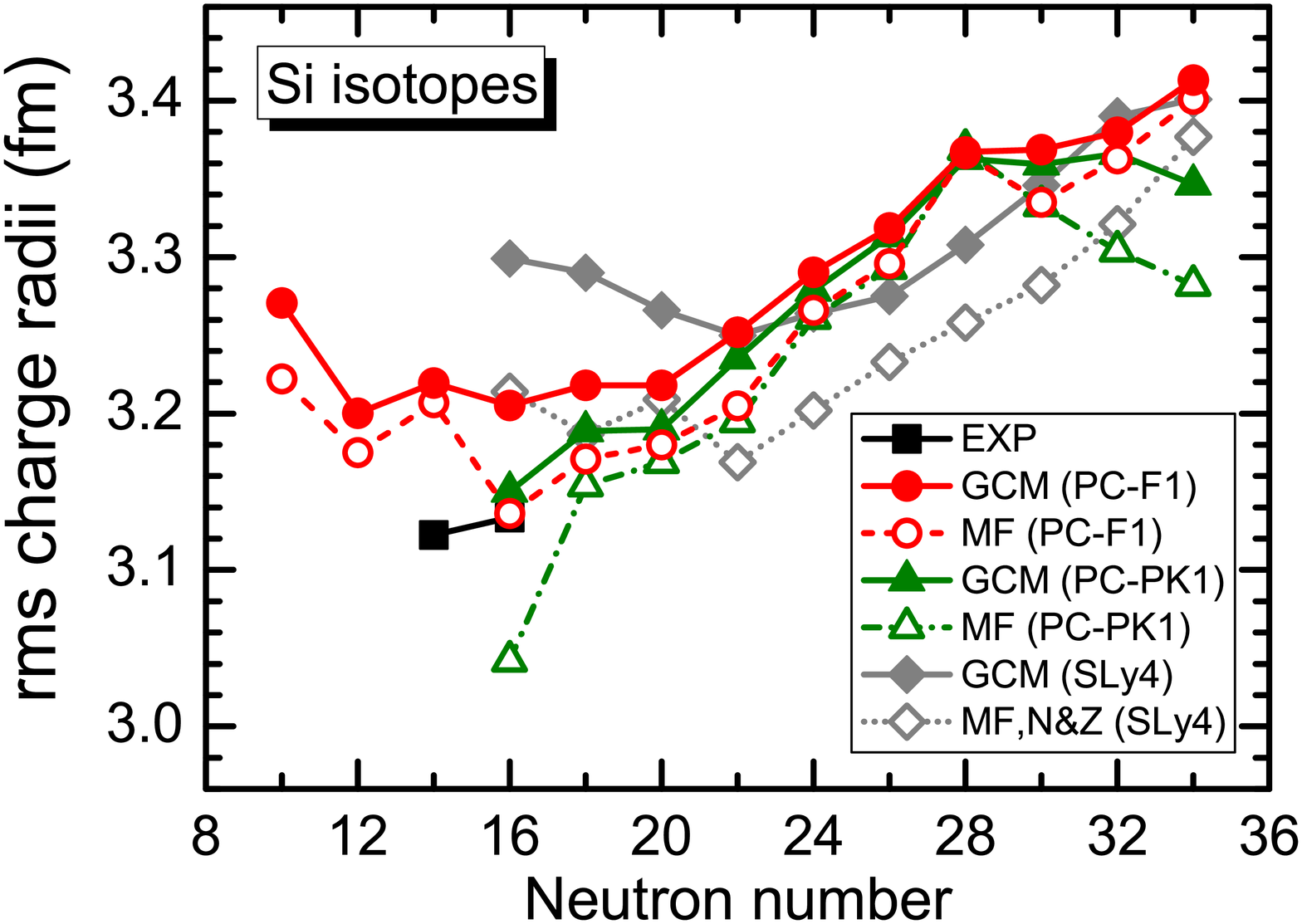}
  \caption{(Color online) Root-mean-square charge radii of Si isotopes from both mean-field and configuration-mixing GCM calculations using the relativistic PC-F1 and PC-PK1 forces as functions
  of neutron number, in comparison with those by the SLy4 force calculations on good particle number ($N$\&$Z$) from Ref.~\cite{Bend06} and available data from Ref.~\cite{Angeli13}.}
  \label{fig:Sirmschar}
  \end{figure}

  It is shown in Fig.~\ref{fig:rmschar} that the systematics of charge radii in $sd$-shell are reproduced rather well. Similar to the binding energies, the beyond-mean-field effect leads to an overestimation of the charge radii for most of the $sd$-shell nuclei, in particular for those around $N=14$, and 16. Dynamic quadrupole deformation effects slightly better the description of the rms charge radius of $^{32}$Mg data than mean-field result. The rms charge radii for Si isotopes calculated by relativistic PC-PK1 force and nonrelativistic SLy4 force~\cite{Bend06} are also given in Fig.~\ref{fig:Sirmschar}. It is a common feature for the three EDFs that the beyond-mean-field calculations overestimate the rms charge radii. Among the three EDFs, the relativistic PC-PK1 force gives the results closest to the data.

\begin{figure}[]
  \centering
  \includegraphics[width=6.5cm]{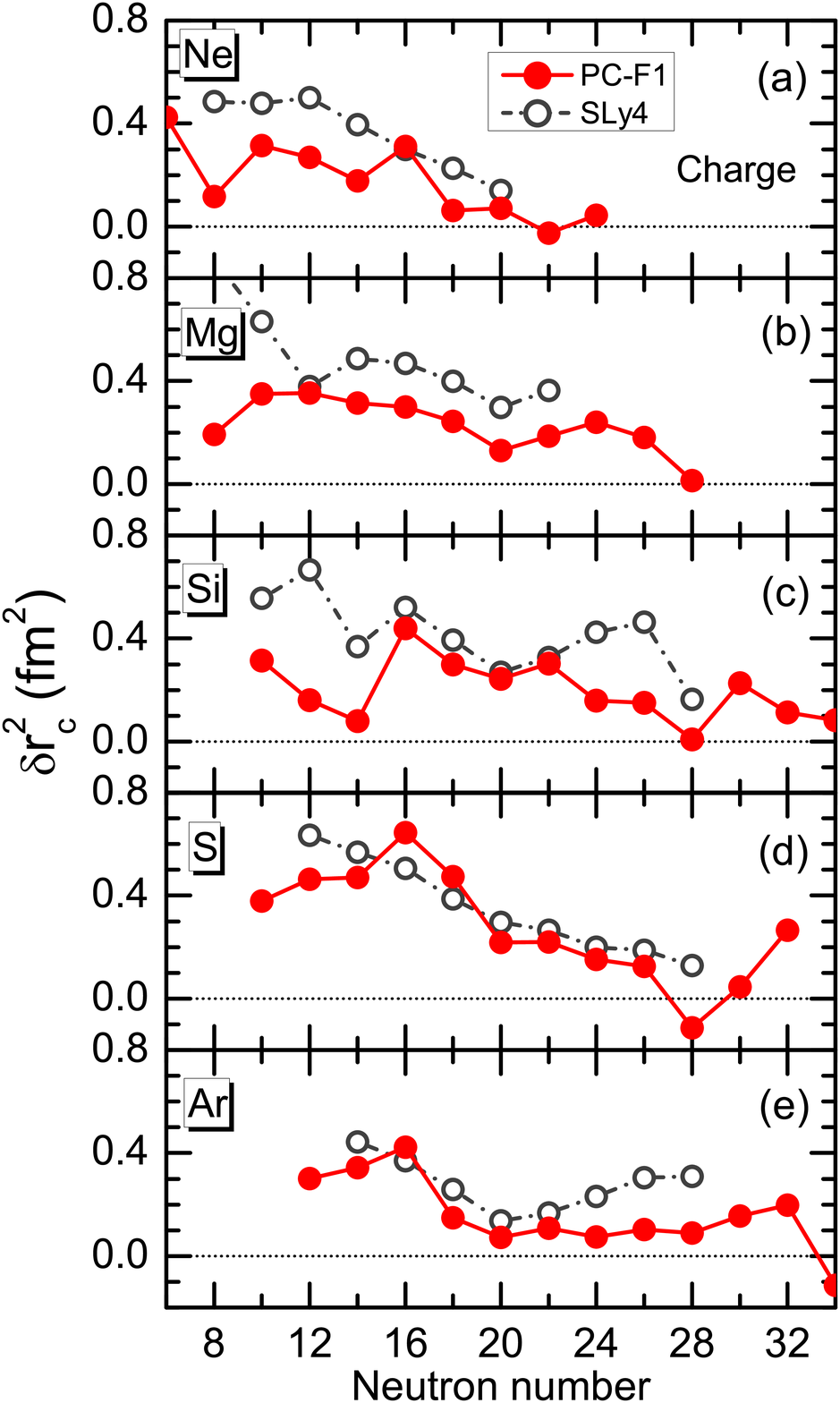}
  \caption{(Color online) Changes $\delta r^2_c$ in the mean-square
  charge radii of $sd$-shell even-even nuclei due to the dynamical correlation effects, where $\delta r^2_c=r^2_c(0^+_1)- r^2_c (\beta_m)$ with $r^2_c(0^+_1)$ and $r^2_c (\beta_m)$ being the mean-square radius of  the $0^+_1$ state and the mean-field ground state, respectively. The $r^2_c (\beta_m)$ by the SLy4 force is the mean-square radius of the minimum of the particle number projected energy curve, while that by the PC-F1 force is the minimum of the mean-field energy curve. All the results obtained by the SLy4 are taken from Ref.~\cite{Bend06}.}
  \label{fig:DCEchar}
  \end{figure}

  The dynamical correlation effects associated with the quadrupole deformation on the charge radii can be seen more clearly in Fig.~\ref{fig:DCEchar}, where the results by the relativistic PC-F1 force are compared with those by the nonrelativistic SLy4 force from Ref.~\cite{Bend06}. The changes $\delta r^2_c$ in the mean-squared charge radii are significant and similar in magnitude for most nuclei by the two totally different forces. For  $^{44}$S and $^{52}$Ar, an opposite contribution to the charge radius is predicted by the relativistic PC-F1 force. It is consistent with the finding in the mean-field calculation with the PC-F1 force that the deformation of ground state is significantly overestimated (cf. Fig.~\ref{fig:sdBE2}), leading to a very large charge radius. The dynamical deformation effect brings the $B(E2;0^+_1\to 2^+_1)$ of $^{44}$S closer to the data.

 \begin{table}
 \caption{The rms deviation of charge radii (in fm) for the 25 even-even $sd$-shell nuclei by the present calculations using the PC-F1 force, in comparison with those by the mean-field HF+BCS calculation using the Skryme SLy4 force~\cite{Bend06} and those by the 5DCH based on the Gogny D1S force~\cite{Delaroche10}. }
 \begin{ruledtabular}
 \begin{tabular}{cc|cc|cc}
% \hline
   \multicolumn{2}{c|}{ PC-F1 } &  \multicolumn{2}{c|}{ SLy4 }    &  \multicolumn{2}{c}{ D1S }  \\
  \hline
   RMF+BCS &   GCM~  & HF+BCS  & GCM~ &   HFB  & 5DCH~    \\
  \hline
   0.032 &   0.057 & 0.033  & 0.094   & 0.035  & 0.061    \\
 \end{tabular}
 \end{ruledtabular}
 \label{rms:charge-radii}
 \end{table}

  Table~\ref{rms:charge-radii} presents the comparison of the rms deviation of charge radii for the 25 even-even $sd$-shell nuclei calculated with both mean-field and beyond-mean-field approaches using three different energy functions. In all the cases, the configuration-mixing beyond-mean-field calculation overestimates the charge radii and thus gives the rms deviation around $0.06\sim0.09$ fm larger than that of the mean-field calculation with rms deviation around $0.03$ fm.

  In recent years, the possible existence of ``bubble" structure in light nuclei has attracted renewed interesting because it may change nuclear shell structure through a weakened spin-orbit interaction~\cite{Grasso09}. The previous studies within the MR-DFT have demonstrated that the ``bubble" structure in light nuclei can be greatly quenched by static and dynamic deformation effects~\cite{Yao12,Yao13,Wu14}.

  \begin{figure}[]
  \centering
  \includegraphics[width=8.5cm]{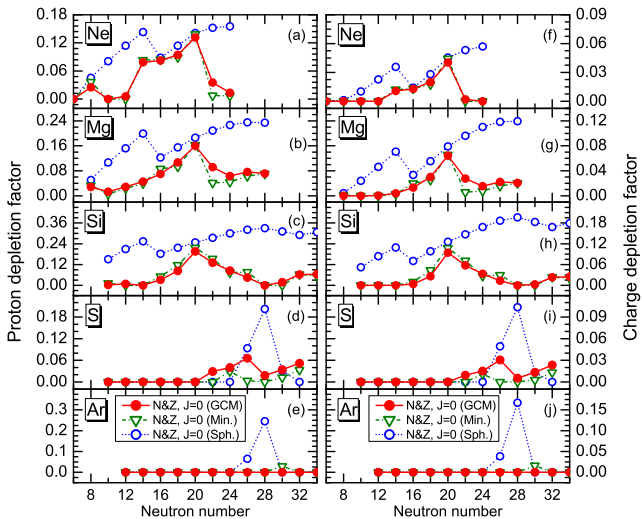}
  \caption{(Color online) Central depletion factor of proton (a-e) and charge (f-j) in $sd$-shell even-even
  isotopes as functions of neutron number. The results are obtained from spherical state (blue line),
  minimum of projected ($N$\&$Z$, $J=0$) energy curve (olive line) and the full configuration mixing (red line) calculation using the PC-F1, respectively. }
  \label{fig:Factor}
  \end{figure}

  In this work, we generalize this investigation for all the $sd$-shell nuclei. Figure~\ref{fig:Factor} displays the central depletion factor in the proton and charge densities of $sd$-shell even-even nuclei, where the depletion factor $F_{\rm max}$ is defined as $F_{\rm max}\equiv(\rho_{\rm max}-\rho_{\rm cent})/\rho_{\rm max}$
  with $\rho_{\rm max}$ being the largest value of the density in coordinate space and $\rho_{\rm cent}$ the value at the
  center $r=0$. In general, one might notice that the depletion factor $F_{\rm max}$ of proton is larger than charge
  depletion factor $F_{\rm max}$ for all the isotopes by a factor of $2\sim3$. It is also seen that in most nuclei the $F_{\rm max}$ is decreased to zero when the dynamical deformation effects are taken into account.  For Ne, Mg, Si isotopes, the depletion factor $F_{\rm max}$ indicates that possible existence of bubble structure in spherical states of $N=14$ and $N=28$ isotones. However, beyond-mean-field calculation shifts the peak position of the depletion factor $F_{\rm max}$ to $N=20$. With the presence of dynamical correlation effects, a noticeable bubble structure remains in $^{30}$Ne, $^{32}$Mg, and $^{34}$Si. In particular, $^{34}$Si is still the best candidate with ``bubble" structure. For sulfur isotopes, the dynamical correlation effects do not always quench the semi-bubble structure, but also help to develop the semi-bubble structure in $^{38,40,46,48}$S. For argon isotopes, the existence of ``bubble" structure is unlikely after  taking into account the dynamical correlation effects~\cite{Wu14}.

  \subsection{$2^+_1$ and $4^+_1$ states}

  \begin{figure}[]
  \centering
  \includegraphics[clip=,width=7cm]{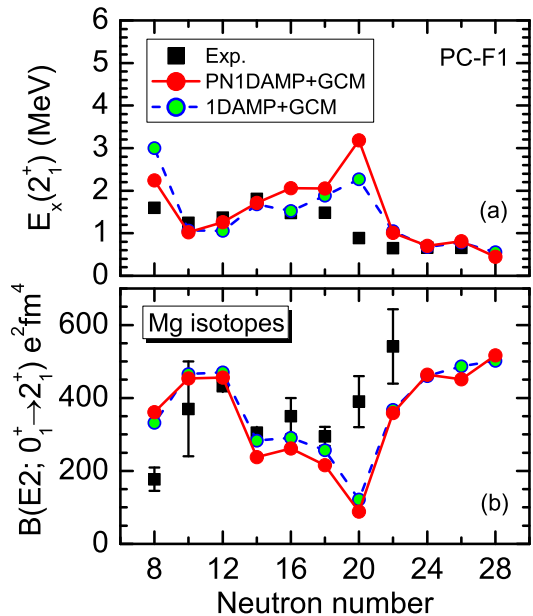}\vspace{-0cm}
  \caption{(Color online) Excitation energies of the $2^+_1$ states (a) and $B(E2; 0^+_1\to 2^+_1)$ values (b) in Mg isotopes as functions of neutron number. The results from PN1DAMP+GCM (solid line) are compared with 1DAMP+GCM
  (dash line)~\cite{Yao11} calculation using the same relativistic density functional PC-F1. Available data~\cite{NNDC}
  are also given for comparison.}
  \label{fig:MgPNP}
  \end{figure}
  Since in the previous work~\cite{Yao11} the properties of the low-lying states of magnesium isotopes have been studied with the PC-F1 force without the particle number projection, we start this section by first examining the effect of particle number projection on these states. PNP before variation can avoid pairing collapse and finally increase the excitation energies of states. Like other multireference density functional methods, the problems of self-interactions and self-pairing might exist in our method. However, we have not seen any evidence that these problems are very serious to be taken into account. We have checked our calculation against the number of mesh points in the Gauge angle and obtained a good plateau condition. All the results are obtained based on this plateau condition. Figure~\ref{fig:MgPNP} shows the comparison of the excitation energy for the $2^+_1$ states and the $B(E2; 0^+_1\to 2^+_1)$ value from the configuration-mixing GCM calculation with and without particle number projection. We remind that in the previous calculation without the  particle number projection, an approximate correction scheme~\cite{Hara82,Bonche90} for particle numbers was implemented, cf. Ref.~\cite{Yao11} for details. It confirms the finding in Ref.~\cite{Yao11} that this correction scheme yields the results very close to those with the exact particle-number projection for all the Mg isotopes. However, the anomalous behavior in $^{32}$Mg is not reproduced in both calculations. As found in Refs.~\cite{Yao09,Yao11}, the onset of large collectivity in the ground state of $^{32}$Mg can be reproduced much better by pumping more nucleons up to the $pf$-shell by adopting a very strong pairing force. With the pairing strengths adjusted to fit the empirical pairing gaps, we obtained the $E2$ transition strength $B(E2; 0^+_1\to 2^+_1)= 313.5$ e$^2$fm$^4$. It gives us a hint that the noncollective configurations of particle-hole excitation across the $N=20$ shell are likely important to reproduce the anomalous behavior in $^{32}$Mg with the relativistic MR-DFT, even though this anomalous behavior was reproduced with the Gogny force~\cite{Rodriguez-Guzman00}. The inclusion of noncollective particle-hole excitation configurations in the model space of our method is in progress.

  \subsubsection{Excitation energy}

  \begin{figure}[]
  \centering
  \includegraphics[width=8cm]{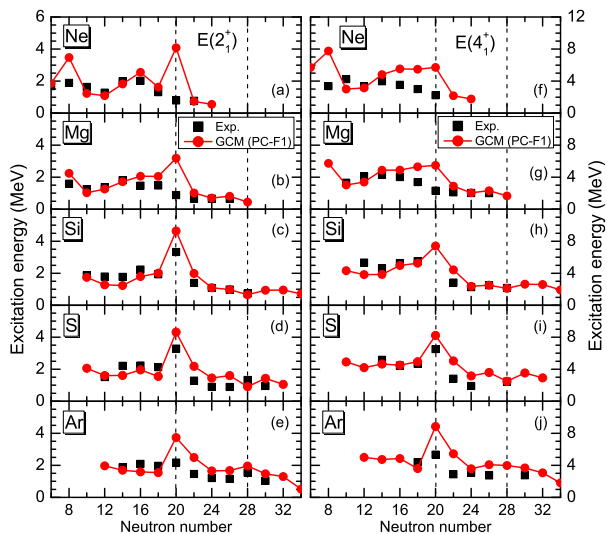}
  \caption{(Color online) Excitation energies of  the $2^+_1$ state (left panel) and
  $4^+_1$ state (right panel) as a function of neutron number in $sd$-shell even-even Ne, Mg,
  Si, S, and Ar isotopes, in comparison with available data from Refs.~\cite{Marinova11,Takeuchi12,Doornenbal13,NNDC}. }
  \label{fig:E24}
  \end{figure}

\begin{figure}[]
  \centering
  \includegraphics[width=6cm]{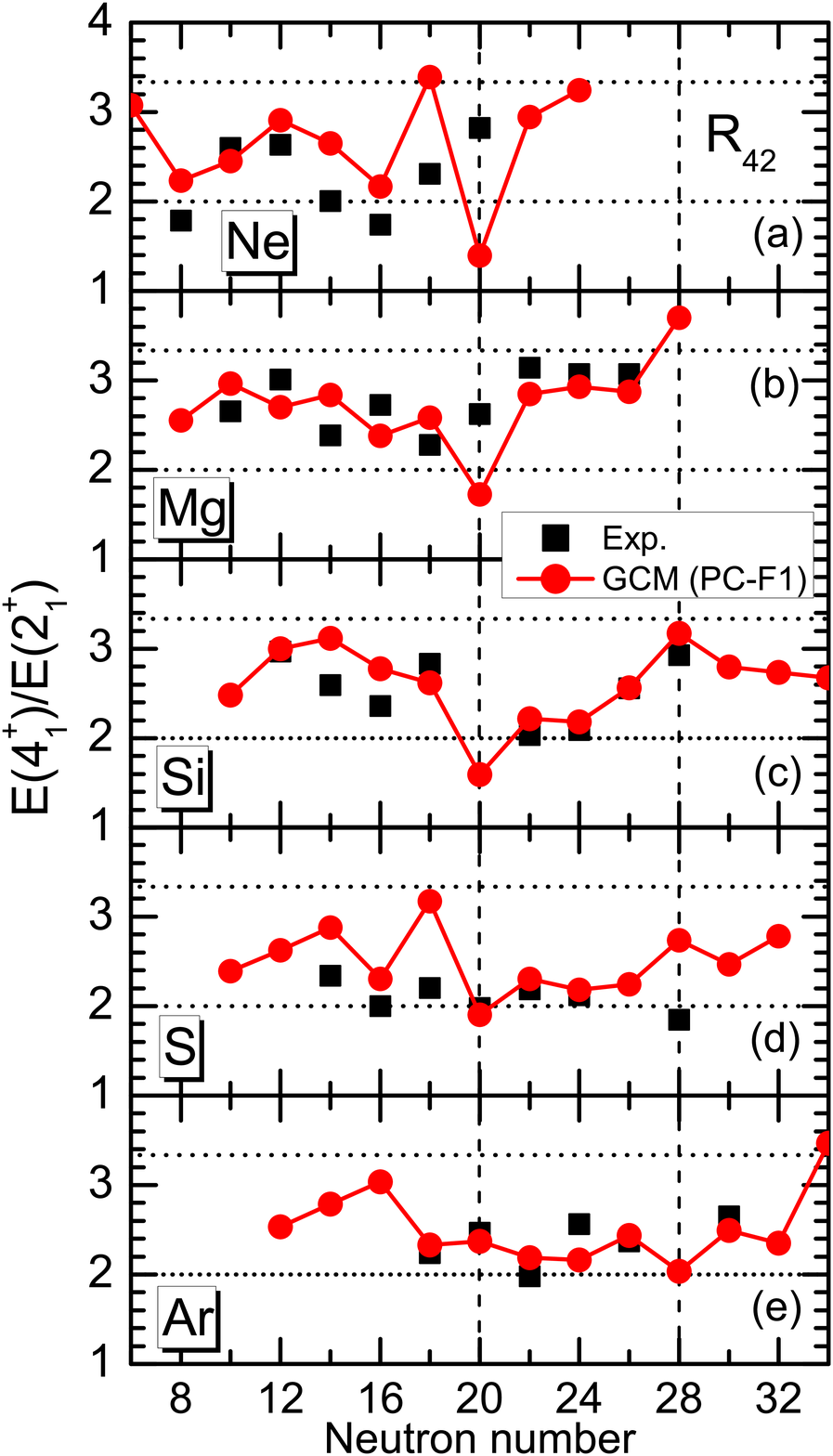}
  \caption{(Color online) Excitation energy ratios $R_{42}=E(4^+_1)/E(2^+_1)$ as a
  function of neutron number in $sd$-shell even-even Ne, Mg, Si, S, and Ar isotopes,
  in comparison with available data from Refs.~\cite{Marinova11,Takeuchi12,Doornenbal13,NNDC}.}
  \label{fig:ratio}
  \end{figure}

  Figure~\ref{fig:E24} shows the calculated excitation energies of the $2^+_1$ and $4^+_1$ states in even-even Ne, Mg, Si, S, and Ar isotopes using the PC-F1 force, in comparison with available data~\cite{Marinova11,Takeuchi12,Doornenbal13,NNDC}. The global behaviors of the excitation energies of $2^+_1$ states are in rather good agreement with data, except for the significant overestimation of the excitation energies in $^{18}$Ne, $^{30}$Ne and $^{32}$Mg. This problem is ascribed to the predicted large $N=8$ and $N=20$ shell gaps in these nuclei. As we mentioned above for $^{32}$Mg, a strong pairing can help to break the robust $N=20$ gap. One might notice that the $4^+_1$ energies are much higher than the data around N=20 in Ne, Mg and Ar.
  We note that pairing collapse is found in most of the nuclei with predicted excitation energies lower than the data. The inclusion of PNP before variation in the mean-field calculation helps to increase the excitation energies. Therefore, even though the effects of triaxiality and time-odd components will lower down the energies, the final results with the PNP before variation can be close to the data, as illustrated in Ref.~\cite{Borrajo15}. The excitation energy ratio $R_{42}\equiv E(4^+_1)/E(2^+_1)$ of the $4^+_1$ state to  $2^+_1$ is displayed in Fig.~\ref{fig:ratio}. For comparison, the limits of a rigid rotor ($3.33$) and a vibrator (2.0) are also plotted.  Most of the data are distributed in the region from 2.0 to 2.5, except for $^{30}$Ne, $^{34, 36, 38}$Mg and $^{42}$Si which are closer to the rotor limit. In the present calculation, we overestimate the ratio $R_{42}$ for $^{28}$Ne, $^{34, 44}$S and underestimate that for the neutron-rich $^{30}$Ne and $^{32}$Mg. Additionally, our calculations show an opposite staggering to the experiment data in $^{22-32}$Mg. Of particular interest is the finding that a rapid increase of $R_{42}$ from $^{38}$Si to $^{42}$Si is shown in both theoretical results and data, exhibiting a transition picture from a spherical vibrator to a rigid rotor. In short, the systematic behavior of $R_{42}$ indicates that the collectivity in most $sd$-shell nuclei below $N=20$ is weak but it starts to develop beyond $N=20$ in neutron-rich side, where the deformation-driving $f_{7/2}$ orbital plays an important role.

  \subsubsection{Electric quadrupole transition and spectroscopic quadrupole moment}

  The electric quadrupole transition strength $B(E2)$ from the initial state
  $(J_i,\alpha_i)$ to the final state $(J_f,\alpha_f)$ is calculated as follows
  \begin{eqnarray}
   &&B(E2; J_i,\alpha_i\rightarrow J_f,\alpha_f)\nonumber\\
     &=& \frac{1}{2J_i+1}
         \Big\vert\sum_{\beta^\prime,\beta}f_{\alpha_f}^{J_{f}\ast}(\beta^\prime)  \langle J_f,\beta^\prime\vert\vert \hat Q_{2}\vert\vert J_i,\beta\rangle
         f_{\alpha_i}^{J_{i}}(\beta)\Big\vert^2\, ,\nonumber\\
  \label{BE2}
  \end{eqnarray}
  where the reduced transition matrix element reads
  \begin{eqnarray}
  &&\langle J_f,\beta^\prime\vert\vert \hat Q_{2}\vert\vert J_i,\beta\rangle\nonumber\\
  &=&\frac{(2J_f+1)(2J_i+1)}{2}
   \sum_{M=-2}^{+2}
  \left(\begin{array}{ccc}
    J_f  &  2     &J_i \\
    0    & M      &-M \\
  \end{array}
  \right) \nonumber\\
  &&  \int_0^\pi d\theta\,\sin(\theta)\,d_{-M0}^{J_i\ast}(\theta)\langle\Phi(\beta^\prime)
  \vert\hat Q_{2M}e^{i\theta\hat J_y}\hat P^N \hat P^Z\vert\Phi(\beta)\rangle \, .\nonumber\\
   \label{Q2}
  \end{eqnarray}

  In the mean-time, we can also calculate the spectroscopic quadrupole moment for each state
   \begin{eqnarray}\label{qspec}
   Q^{\rm spec}(J_\alpha^\pi)
  % &=& \sqrt{\frac{16\pi}{5}}\langle\Psi_\alpha^{J}\vert
  %  \hat Q_{20}\vert\Psi_\alpha^{J}\rangle\nonumber\\
  &=& \sqrt{\frac{16\pi}{5}}
  \left(\begin{array}{ccc}
     J  &  2         &J \\
     J & 0      &-J \\
  \end{array}
  \right)\sum_{\beta,\beta^\prime}f_\alpha^{J\ast}(\beta^\prime) \nonumber\\
  &&\times \langle J,\beta^\prime
  \vert\vert\hat Q_2\vert\vert J,\beta\rangle f_\alpha^{J}(\beta) \, .\nonumber\\
  \end{eqnarray}
  where $\hat Q_{2M}\equiv er^2Y_{2M}$ is the electric quadrupole moment operator.
  Since the $B(E2)$ values and spectroscopic quadrupole moments $Q^{\texttt{spec}}(J,\alpha)$ are
  calculated in the full configuration space, there is no need to introduce effective charge,
  and $e$ simply corresponds to bare value of the proton charge.

  \begin{figure}[]
  \centering
  \includegraphics[width=6.5cm]{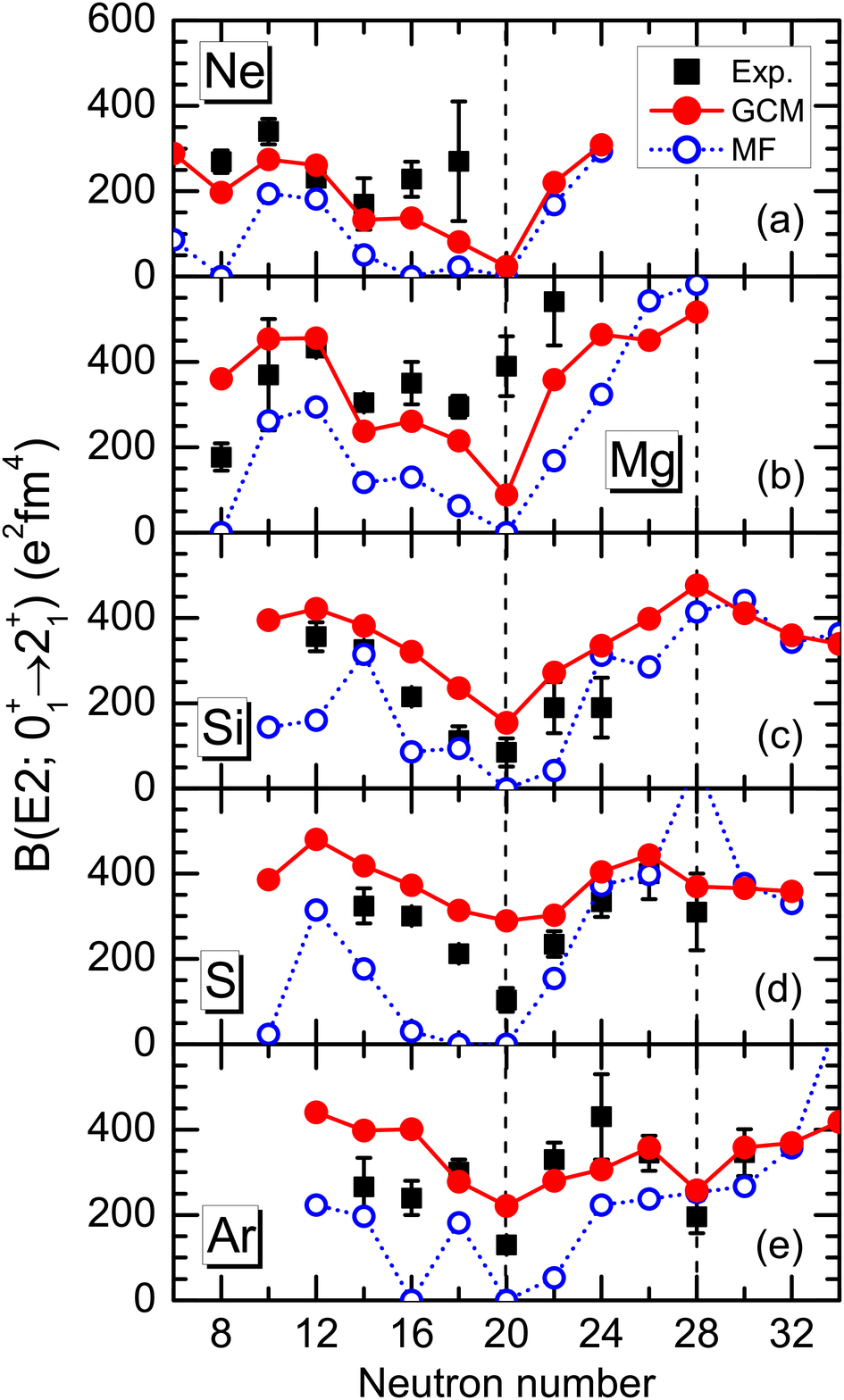}
  \caption{(Color online) Electric quadrupole transition strengths $B(E2; 0^+_1 \to 2^+_1)$ as
  functions of the neutron number in $sd$-shell even-even isotopes. The results obtained from BMF (solid line)
  are compared with those values derived from the deformation $\beta_m$ predicted by MF (dash line)
  using the same relativistic density functional PC-F1. Data are taken from Refs.~\cite{NNDC,Pritychenko14}. More
  details see the text.}
  \label{fig:sdBE2}
  \end{figure}

  Figure~\ref{fig:sdBE2} displays the calculated $B(E2; 0^+_1 \to 2^+_1)$ values in $sd$-shell nuclei with configuration-mixing GCM using the PC-F1 force as functions of neutron number. For comparison, we also plot the mean-field results derived from the deformation $\beta_m$ of the minimum of mean-field energy curve using the formula of the rigid-rotor model~\cite{Ring80}
  \beq
  B(E2; 0^+_1 \to 2^+_1)=\Big(\dfrac{3ZeR^2}{4\pi}\Big)^2\beta^2_m\, .
  \eeq

  Coincident with the systematic behavior of the excitation energy of $2^+_1$ state in Fig.~\ref{fig:E24}, a reasonable good agreement with data is observed in  the $B(E2; 0^+_1 \to 2^+_1)$ values except for again the underestimation of transition strengths in nuclei around $^{32}$Mg and an overestimation of those values in $^{36}$S and $^{38}$Ar. In particular, Fig.~\ref{fig:sdBE2} shows that the results of the present beyond-mean-field calculation are much closer to the data than the mean-field predictions. The mean-field calculation underestimates the $B(E2; 0^+_1 \to 2^+_1)$ value systematically, except for the significant overestimation in $^{44}$S and $^{52}$Ar. This overestimation has already been shown in their charge radii which can be reduced to close to data by the beyond-mean-field effect and become closer to the data as illustrated in Fig.~\ref{fig:DCEchar}.

  \begin{figure}[]
  \centering
  \includegraphics[width=6.5cm]{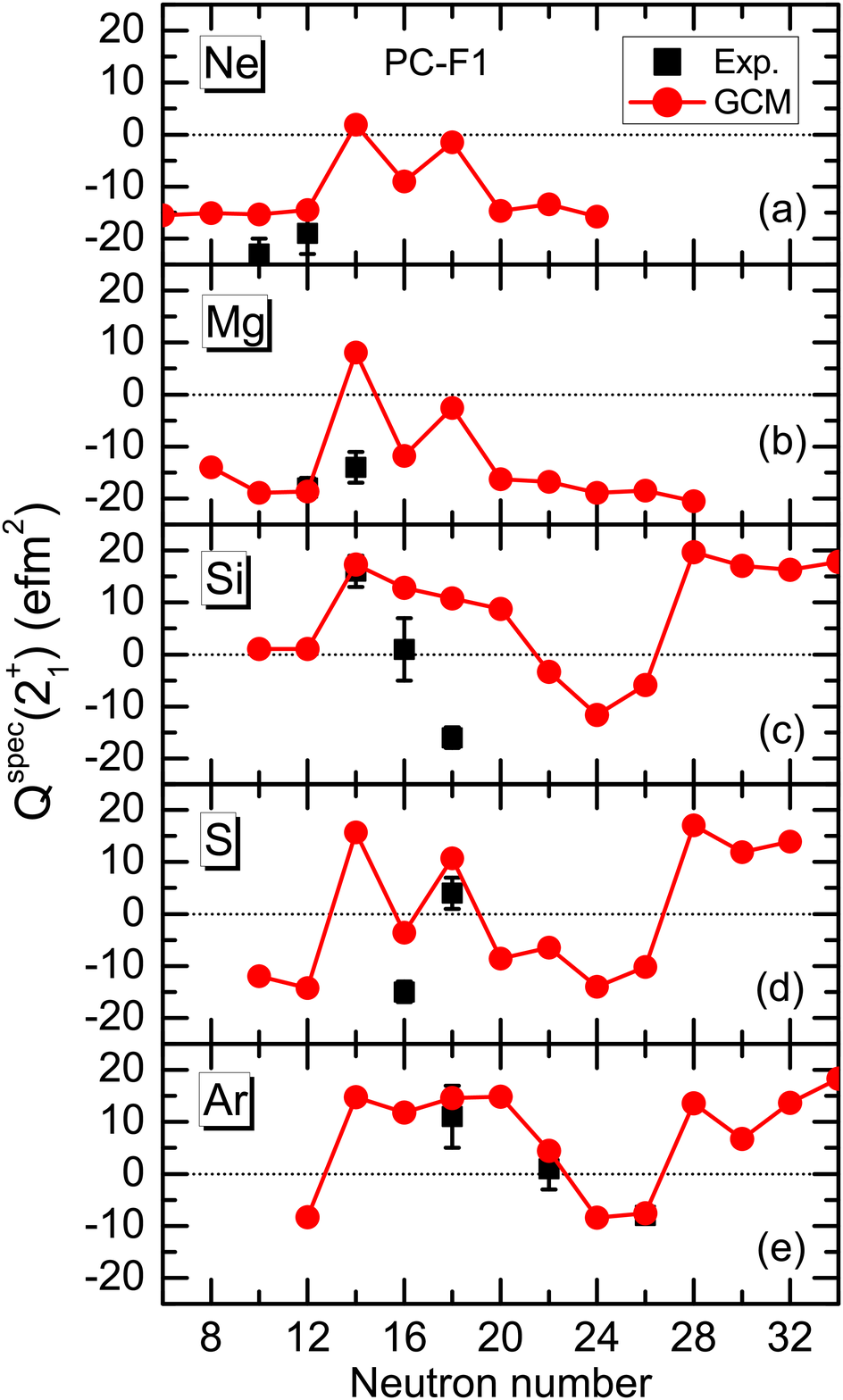}
  \caption{(Color online) Spectroscopic quadrupole moments $Q^{\texttt{spec}}$ of the first $2^+$ as a function of neutron number in $sd$-shell even-even isotopes, in comparison with available data from Ref.~\cite{Stone14}.}
  \label{fig:Qspec}
  \end{figure}
  The information of the dominated shape in the $2^+_1$ state can be learnt from its spectroscopic quadrupole moment  $Q^{\texttt{spec}}$ which is plotted in Fig.~\ref{fig:Qspec}. For a single-shape dominated state, a positive or negative value of $Q^{\texttt{spec}}$ indicates the nucleus with oblate or prolate shape, respectively. However, for the case with a very small $Q^{\texttt{spec}}$, it is difficult to distinguish if the state is almost spherical or with equally weighted oblate and prolate shapes or with $\gamma$-soft character. The information of excitation energy and $E2$ transition strength becomes very important in this case. For instance, $^{30}$Si and $^{40}$Ar have almost zero $Q^{\texttt{spec}}$ for the $2^+_1$ state (see Fig.~\ref{fig:Qspec}) but actually with very different structure. The information of the ratio $R_{42}$ and $B(E2; 0^+_1 \to 2^+_1)$, together with mean-field energy surface (see also Ref.~\cite{CEA}), indicates that $^{30}$Si is oblate deformed with $\gamma$-soft character, while $^{40}$Ar is a rather good vibrator.

  It is shown in Fig.~\ref{fig:Qspec} that the systematics of the calculated $Q^{\texttt{spec}}$ of $2^+_1$ states in Ne and Mg isotopes are similar and those in Si, S and Ar isotopes are similar. Most of the Ne and Mg isotopes are predicted to be prolate deformed except for $^{24,28}$Ne and $^{26,30}$Mg, all of which have transitional characters. For the $\gamma$-soft nucleus $^{26}$Mg, it has already been found in Ref.~\cite{Yao11} that both the Gogny D1S force  and relativistic PC-F1 force predict a positive $Q^{\texttt{spec}}$ value for the $2^+_1$ (opposite to the sign of data) and a negative value for the $4^+_1$ state. The $2^+_1$ state of $^{30}$Si is also predicted to have dominate oblate shape by both forces (see Ref.~\cite{CEA} for the energy surface by the Gogny force), in contrary to the data. Therefore, new  measurements on the spectroscopic quadrupole moments of the $2^+_1$ state in $^{26}$Mg and $^{30}$Si with Coulomb excitations are suggested to confirm the data.

  \begin{figure}[]
  \centering
  \includegraphics[width=8.5cm]{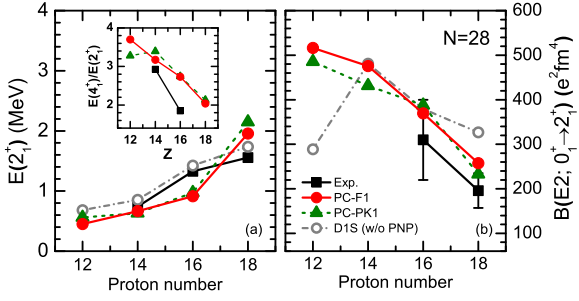}
  \caption{(Color online) (a) Excitation energy $E_x(2^+_1)$ and (b) $B(E2; 0^+_1\to 2^+_1)$ value as functions of proton number in $N=28$ isotones. The results of the Gogny D1S from Ref.~\cite{Rodriguez-Guzman00} were obtained without PNP. The inset panel shows the ratio $R_{42}$. Data are taken from Ref.~\cite{NNDC}.}
  \label{fig:N=28}
  \end{figure}

  Figure~\ref{fig:N=28} shows the calculated excitation energy $E_x(2^+_1)$ and the $B(E2; 0^+_1\to 2^+_1)$ value in $N=28$ isotones with three different forces, in comparison with available data. The results by the Gogny D1S force were obtained in Ref.~\cite{Rodriguez-Guzman00} without particle number projection (PNP). All the three forces  calculations
  reproduce the systematics of available data. However, the $B(E2; 0^+_1\to 2^+_1)$  value of $^{40}$Mg by the D1S force is only about half of the results by the two relativistic forces. Therefore a measurement on this quantity in $^{40}$Mg will provide a stringent test of nuclear EDFs.

  To study the relative contribution of neutrons and protons to the nuclear multipole ($J$) excitations from ground state ($0^+_1$) to excited state ($J^+_\alpha$), we introduce a neutron-proton decoupling factor
  \begin{equation}\label{eqeta}
    \eta=\frac{M^J_n/M^J_p}{N/Z} \, ,
  \end{equation}
  where $M^J_n$ and $M^J_p$ are the multipole transition matrix elements of neutrons and protons, respectively
   \beqn
   \label{TME}
   M^{J}
   =\int^\infty_0 \! dr \, r^{J+2} \, \rho^{J\alpha}_{01, J}(r)\, ,
   \eeqn
   calculated with the reduced transition density of neutrons and protons~\cite{Yao15}
    \beqn
   \rho^{J\alpha}_{01, J}(r)
  &=&\sqrt{(2J+1)} \sum_{\beta'\beta} f^{J\ast}_{\alpha}(\beta')  f^{0}_{1}(\beta) \nonumber\\
   &&\times \int d\hat{\br}  \rho^{J00}_{\beta'\beta}(\br) Y_{J0}(\hat{\br})\, .
  \eeqn
   where the $\rho^{JK0}_{\beta'\beta}(\br)$ with $K=0$ has been given in Eq.~(\ref{rho_JKNZ}). We note that if the neutrons and protons have the same quadrupole deformations, the factor $\eta$ should be one according to the rigid rotor model. Any deviation from one indicates the decoupling of neutrons and protons.

  \begin{figure}[]
  \centering
  \includegraphics[width=6.5cm]{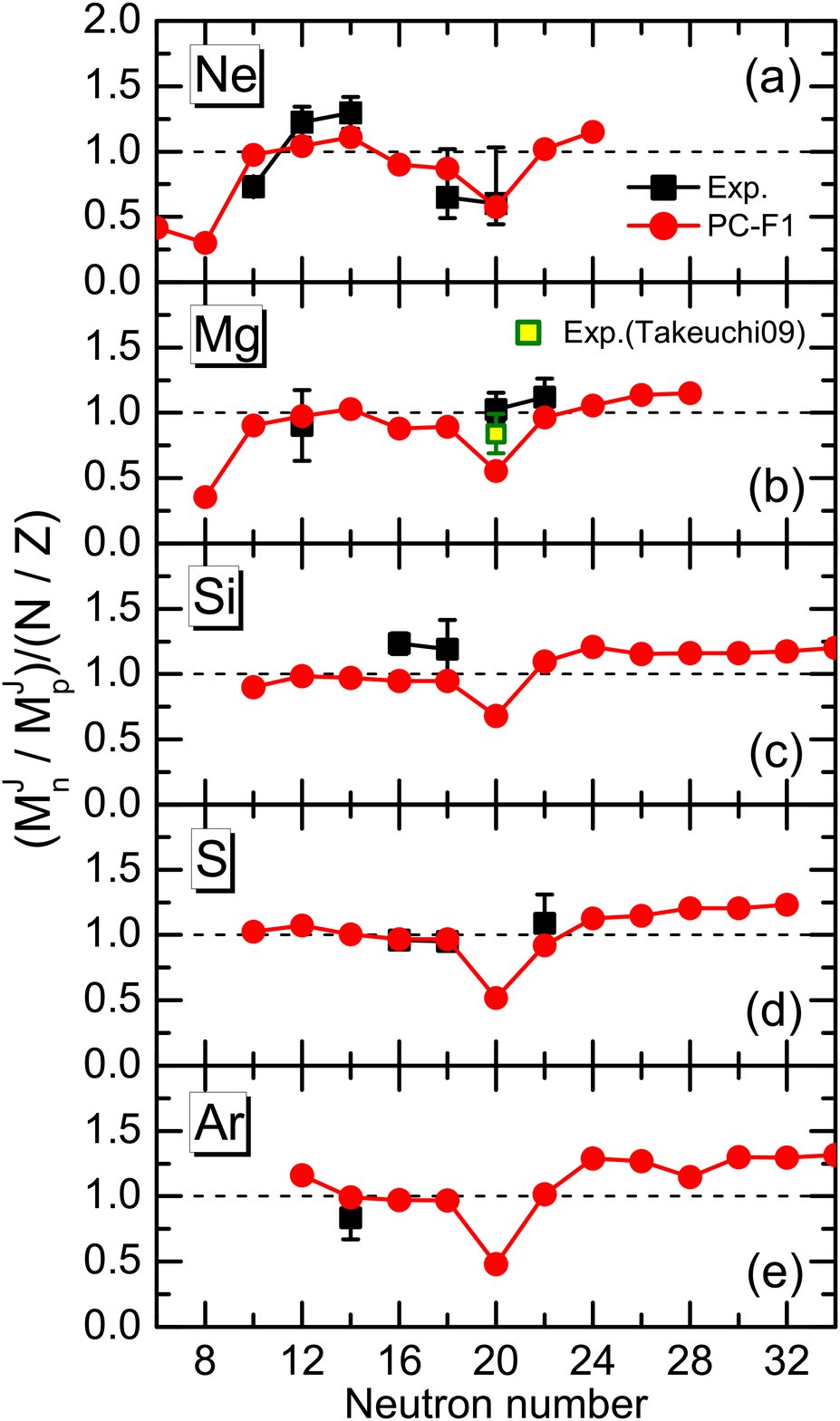}
  \caption{(Color online) Neutron-proton decoupling factor $\eta$ [cf. Eq. (\ref{eqeta})] of the $2_1^+$ state of
   even-even $sd$-shell nuclei as a function of neutron number, in comparison with available data
   from Refs.~\cite{ALEXANDER82,Khandaker91,Kennedy92,Kelley97,Alamanos99,Cottle02,Takeuchi09}. The experimental
   data for $^{28,30}$Ne and $^{32,34}$Mg are extracted using the formula and corresponding parameters
   given in Ref.~\cite{Michimasa14}.}
  \label{fig:ratioeta}
  \end{figure}

  Figure~\ref{fig:ratioeta} displays  the neutron-proton decoupling factor $\eta$ for the $2^+_1$ in  $sd$-shell nuclei, in comparison with available values extracted  from  the data of $(p, p')$ inelastic scattering  and the data of $B(E2; 0^+_1 \to 2^+_1)$. The available data are reproduced rather well by the calculation using the PC-F1 force. The theoretical results show that the protons contribute much more to the $E2$ excitation to the $2^+_1$ than the neutrons in the nuclei with neutron numbers at $N=8$ and $20$. However, for the neutron-rich $N=28$ isotones, neutrons play a dominant role in the $E2$ transitions, indicating again the erosion of the $N=28$ shell gap. For the stable nuclei with $N\simeq Z$, decoupling factor $\eta$ is close to one, indicating the similar quadrupole deformations of neutrons and protons  in their ground states.

  The distribution of the electric multipole transition strength in coordinate space from ground state ($0^+_1$) to excited state ($J^+_\alpha$) can be learnt from the measured Coulomb form factor in electron inelastic scattering
  \beqn
  \label{formfactor}
  F_J(q) =\frac{\sqrt{4\pi}}{Z}\int_0^\infty   dr r^2 \rho^{J\alpha}_{01, J}(r) j_J (qr)\, ,
  \eeqn
  where the coefficient $\sqrt{4\pi}/Z$ is chosen so that the elastic form factor $F_0(q)$ is unity at $q = 0$.

  \begin{figure}[]
  \centering
  \includegraphics[width=8cm]{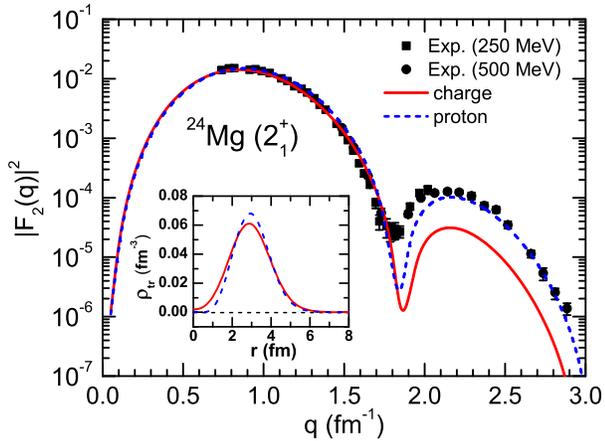}
  \caption{(Color online) Longitudinal C2 form factor $|F_2(q)|^2$ for the transition from the
  ground state $0^+_1$ to the excited state $2^+_1$ for ${}^{24}$Mg. The inset panel shows the
  corresponding transition densities. Data are taken from Ref.~\cite{Li74} (squares and circles).}
  \label{fig:Mg24F2}
  \end{figure}

  \begin{figure}[]
  \centering
  \includegraphics[width=8cm]{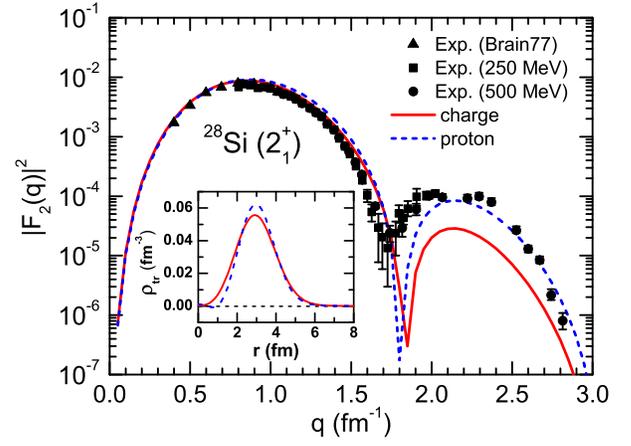}
  \caption{(Color online) Same as Fig.~\ref{fig:Mg24F2}, but for $^{28}$Si. Data  are taken from Refs.~\cite{Li74,Brain77}.}
  \label{fig:Si28F2}
  \end{figure}

\begin{figure}[]
  \centering
  \includegraphics[width=8cm]{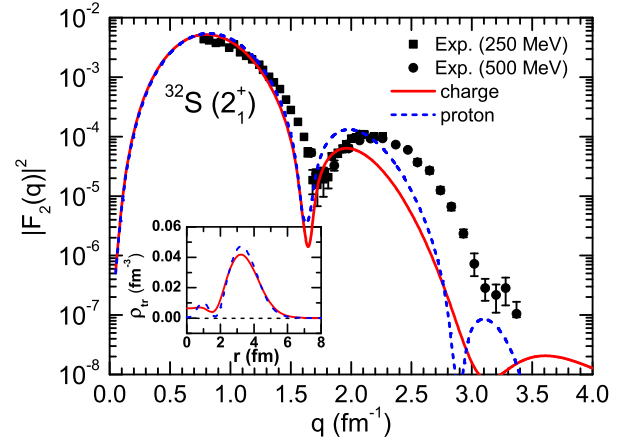}
  \caption{(Color online) Same as Fig.~\ref{fig:Mg24F2}, but for $^{32}$S.}
  \label{fig:S32F2}
  \end{figure}

  Taking ${}^{24}$Mg, ${}^{28}$Si and ${}^{32}$S as examples, we plot the inelastic form factors from the ground state $0^+_1$ to
  the $2^+_1$ state in Fig.~\ref{fig:Mg24F2}, Fig.~\ref{fig:Si28F2} and \ref{fig:S32F2}, respectively. The theoretical results calculated with both proton and charge transition density are plotted. It is shown in the three nuclei that the form factors at the momentum transfer $q$ below the first minimum are reproduced rather well. However, the high-$q$ inelastic form factors are significantly underestimated after the finite-size effect of proton is taken into account. This problem is consistent with the overestimated charge radii by the configuration-mixing calculation, as shown in Fig.~\ref{fig:rmschar}. Similar phenomenon has already been found in the previous works~\cite{Yao15,Mei15-2}.

 \subsubsection{Global analysis}

  \begin{figure}[]
  \centering
  \includegraphics[width=6cm]{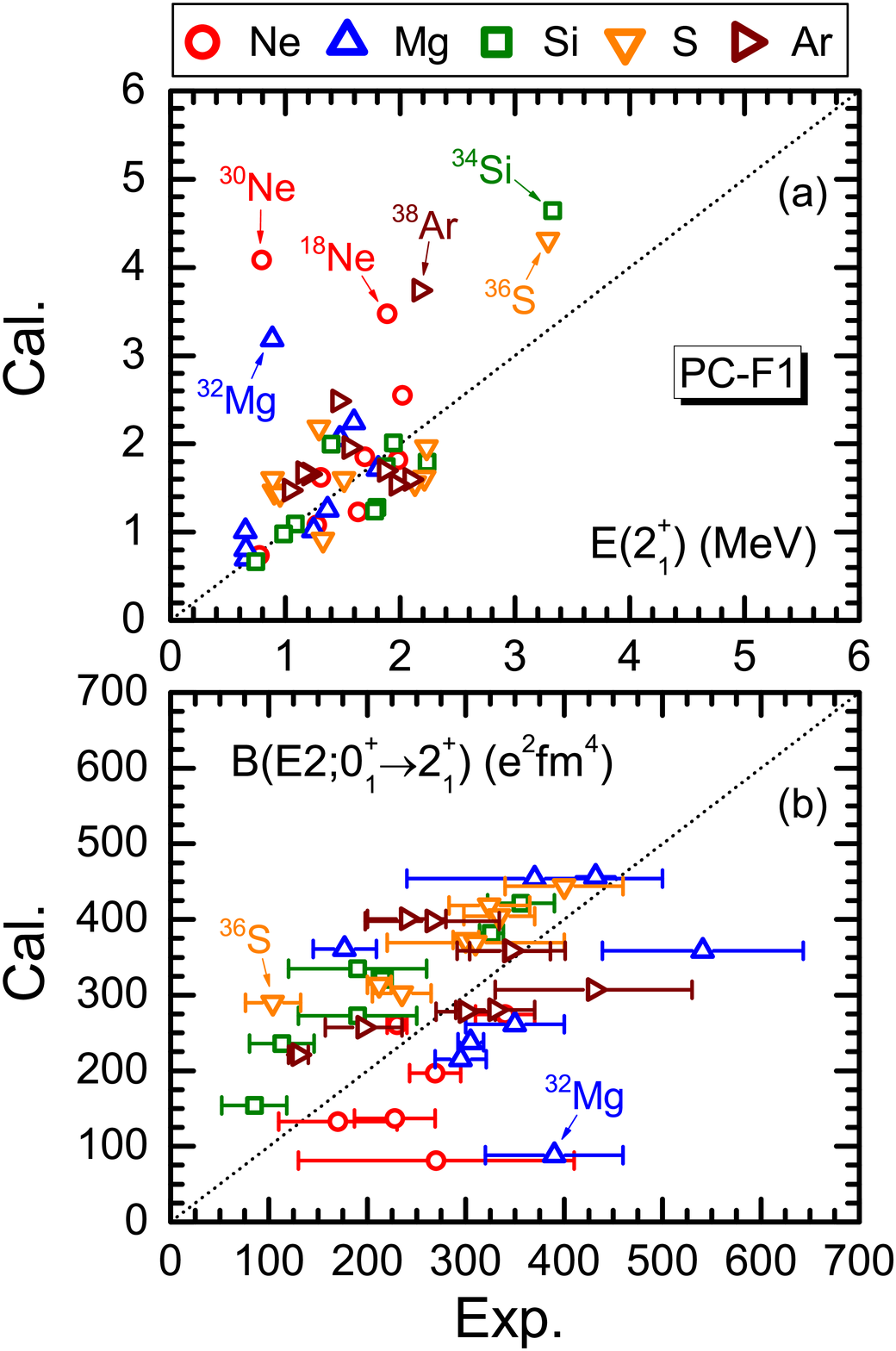}
  \caption{(Color online) (a) Calculated $E(2^+_1)$ excitation energies of 48 even-even nuclei in
  $sd$-shell region as a function of their experimental values. (b) Calculated $B(E2; 0^+_1 \to 2^+_1)$
  transition strengths of 38 even-even nuclei in $sd$-shell region as a function of their experimental
  values. Experimental data are taken from Ref.~\cite{NNDC}.}
  \label{fig:globalcomp}
  \end{figure}

  In this section, we examine the global performance of the MR-DFT for the excitation energy of $2^+_1$ state and the  $B(E2; 0^+_1 \to 2^+_1)$ value in $sd$-shell nuclei.  Figure~\ref{fig:globalcomp} displays the calculated values against the corresponding data.  The excitation energies are mainly distributed along the diagonal line, except for again for $^{18}$Ne and the $N=20$ isotones $^{30}$Ne,  $^{32}$Mg,  $^{34}$Si, and $^{36}$S. Taking into account the error bars, the $B(E2)$ values are much better reproduced than the $E(2^+_1)$ except for $^{32}$Mg and $^{36}$S. We calculate the relative rms error $\sqrt{\frac{1}{N}\sum\limits_{i}^N\left(\frac{O^{cal.}_i-O^{exp.}_i}{O^{exp.}_i}\right)^2}$ and  obtain the value $\simeq$ 0.79 and 0.54  for the set of 48 excitation energies of $2^+_1$ states and 38 transition strengths of $B(E2)$ values, respectively. By excluding the $^{30}$Ne and $^{32}$Mg, this value is reduced to $\simeq$ 0.36 for the excitation energy. After  taking into account the experimental error bars for the $B(E2)$ values, the relative rms error becomes $\simeq$ 0.34.

  Moreover, we follow Refs.~\cite{Sabbey07,Bertsch07} to compare theoretical results $O^{cal.}$ and experimental data $O^{exp.}$ on a logarithmic scale
  \beq
  R=\log(O^{cal.}/O^{exp.})\, .
  \label{loger}
  \eeq

\begin{figure}[]
  \centering
  \includegraphics[width=6cm]{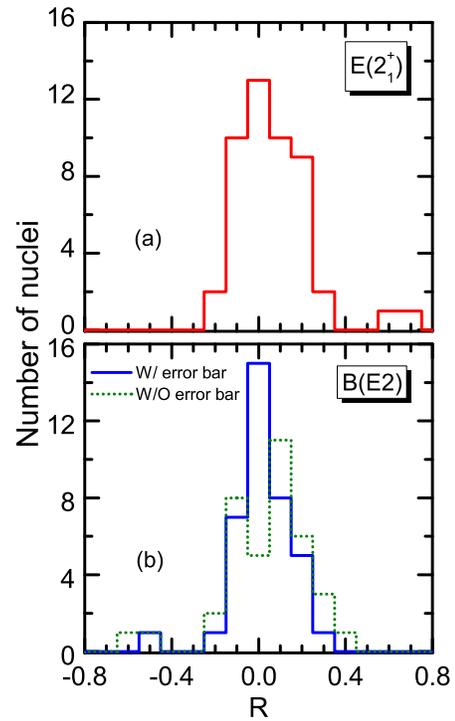}
  \caption{(Color online) Histogram of the logarithmic errors $R$ defined in Eq. (\ref{loger}).
  (a) Top panel: Excitation energies of the first $2^+$ states for 48 even-even nuclei.
  (b) Bottom panel: $B(E2; 0^+_1 \to 2^+_1)$ transition strengths of 38 even-even nuclei in $sd$-shell region.
  The distribution of $B(E2)$ without experimental error bar is also given (dash line).
  Experimental data are taken from Ref.~\cite{NNDC}. }
  \label{fig:histogram}
  \end{figure}

  The width of the distribution of $R$ provides an important indicator on the accuracy and reliability of the calculation. Figure~\ref{fig:histogram} shows the histogram of the logarithmic errors $R$ for both the excitation energies of $2^+_1$ state and the transition strengths of $B(E2; 0^+_1\to 2^+_1)$ values.

  The logarithmic errors $R$ of the excitation energies slightly tend to positive values, indicating a soft
  overestimation of the energy by our calculation. The large values of $R=0.6$ and 0.7 are from $^{32}$Mg
  and $^{30}$Ne, respectively.

  For the set of 48 even-even nuclei in $sd$-shell, the average logarithmic
  error is $\langle R \rangle\approx 0.07$ but the averaged absolute value of the error is much larger
  with $\langle|R|\rangle \approx 0.14$. We also calculate the root-mean-square (rms) deviation, i.e., the dispersion around the average is found to be $\langle(R-\langle R\rangle)^2\rangle^{1/2}\approx 0.17$.

  The logarithmic errors $R$ of the $B(E2)$ values are plotted in Fig.~\ref{fig:histogram} (b). There is a hollow at $R=0.0$ and a peak at $R=0.1$ when the error bars of experimental data are not taken into account (dash line). The values of $R=-0.5$ and $-0.6$ are from $^{28}$Ne and $^{32}$Mg, respectively. However, after taking the experimental error bars into account, the logarithmic errors $R=0$ are moved to be closer to zero. For the set of 38 even-even nuclei, the average error is found to be $\langle R \rangle\approx 0.02$ but the averaged absolute value and rms deviation error are $\langle|R|\rangle \approx 0.09$ and $\langle(R-\langle R\rangle)^2\rangle^{1/2}\approx 0.14$, respectively. These values are smaller than that for the excitation energies of the $2^+_1$ states. It shows again that the $B(E2)$ transition strengths are better reproduced than the excitation energies. The detailed values are given in Table~\ref{tab5}.

  \begin{table}
  \caption{Statistics for the performance of the PN1DAMP+GCM model with PC-F1 force for different set of observables. Two nuclei ($^{30}$Ne, $^{32}$Mg) are excluded in set II. The label ``w/o'' and ``w/'' represents the calculation without and with taking into account the error bars in the data.}
  \tabcolsep=5pt
  \begin{tabular}{ccccccccc}
  \hline\hline
  \multicolumn{1}{c}{  }& \multicolumn{1}{c}{Nuclei}  &  \multicolumn{1}{c}{$\langle R \rangle$ }
  & \multicolumn{1}{c}{$\langle |R| \rangle$}  & \multicolumn{1}{c}{$\sqrt{\langle(R-\langle R\rangle)^2\rangle}$}\\
  \hline
  E($2_1^+$) (I)                &~  $48$  &~ $0.07$   &~ $0.14$ &~  $0.17$ \\
  E($2_1^+$) (II)               &~  $46$    &~ $0.04$   &~ $0.11$ &~  $0.13$ \\
  $B(E2)$ (w/o)     &~  $38$  &~ $0.03$ &~ $0.16$  &~ $0.21$  \\
  $B(E2)$ (w/)      &~  $38$  &~ $0.02$ &~ $0.09$  &~ $0.14$  \\
  \hline \hline
  \end{tabular}
  \label{tab5}
  \end{table}

%%%%%%%%%%%%%%%%%%%%%%%%%%%%%%%%%%%%%%%%%%
%
\section{summary}\label{Sec.IV}
%
%%%%%%%%%%%%%%%%%%%%%%%%%%%%%%%%%%%%%%%%%%

  We have carried out a comprehensive study of the ground state and the $2^+_1, 4^+_1$ states in $sd$-shell nuclei with the multireference density functional theory based on relativistic point-coupling energy density functionals. The global performance of this beyond-mean-field approach has been discussed in comparison with previous similar calculations based on nonrelativistic energy density functionals and with available data. Our findings are summarized as follows:
  \begin{itemize}
  \item The formula based on cranking approximation for the dynamic correlation energy turns out to be a good approximation for open-shell nuclei, but not for weakly-deformed or spherical nuclei.
  \item For the relativistic PC-F1 force, the beyond-mean-field effects increase the rms deviation of binding energies and charge radii from 2.94 MeV to 3.61 MeV for 56 $sd$-shell nuclei, and from
        0.032 fm to 0.057 fm for 25 $sd$-shell nuclei, respectively, but decreases the rms deviation for two-neutron separation energies from 2.07 MeV to 1.53 MeV for 51 $sd$-shell nuclei. This phenomenon turns out to be a common feature for the energy functionals adjusted at the mean-field level.
  \item The static- and dynamic-deformation effects smooth the density in the interior region and smear the density around the surface. As a consequence, the central depletion in density is
        overall quenched significantly, while the rms charge radii are systematically overestimated. Among the $sd$-shell nuclei, $^{34}$Si remains the best candidate with``bubble'' structure.
  \item The transition strength $B(E2;0^+_1 \to 2^+_1)$ values are much better described by the GCM calculations than the mean-field
        calculations. The latter underestimates the $B(E2)$ values systematically due to deformation collapse or softness of energy surface.
  \item Large neutron-proton decoupling phenomenon is found in $^{16,18}$Ne, $^{20}$Mg and $N=20$ isotones.
  \item The longitudinal Coulomb (CL) $C_2$ form factor $F_2(q)$ is calculated and the experiment data are reproduced rather well at the low momentum transfer $q$ values for the $N=Z$ nuclei
        $^{24}$Mg, $^{28}$Si and $^{32}$S. The form factors at high-$q$ region are underestimated due to too much mixing of large deformed configurations, leading to an enhanced $E2$ transition in these nuclei.
  \end{itemize}

  Finally, we note that in the present study, the configurations are limited to have time-reversal invariant axially deformed mean-field states. To further improve the description of the low-lying states, one of the choices is to extend the model space by including noncollective configurations with particle-hole excitations or the states cranked to different frequency. Moreover  an energy density functional that is parameterized at the beyond-mean-field level is highly required to achieve a better agreement with the  experimental data. This is true in particular for the ground-state properties. Works along these directions are in progress.

%%%%%%%%%%%%%%%%%%%%%%%%%%%%%%%%%%%%%%%%%%

  \begin{acknowledgements}
  We thank J. M. Yao for critical reading of the manuscript and thank Z. P. Li and J. Meng, T. R. Rodr\'{i}guez for valuable  discussions. This work was supported in part by the NSFC under Grants No. 11335002, No. 11475140, and No. 11275160, the Research Fund for the Doctoral Program of Higher Education under Grant No. 20110001110087.
  \end{acknowledgements}


\begin{thebibliography}{99}

 \bibitem{Mueller93} A. C. Mueller and B. M. Sherrill,
                     Ann. Rev. Nucl. Part. Sci. \textbf{43}, 529-583 (1993).
 \bibitem{Tanihata95} I. Tanihata, Prog. Part. Nucl. Phys. \textbf{35}, 505-573 (1995).
 \bibitem{Hansen95} P. G. Hansen, A. S. Jensen, and B. Jonson,
                    Ann. Rev. Nucl. Part. Sci. \textbf{45}, 591-634 (1995).
 \bibitem{Jonson04} B. Jonson, Phys. Rep. \textbf{389}, 1-59 (2004).
 \bibitem{Jensen04} A. S. Jensen, K. Riisager, D. V. Fedorov, and E. Garrido,
                      Rev. Mod. Phys. \textbf{76}, 215-262 (2004).
 \bibitem{Motobayashi95} T. Motobayashi, Y. Ikeda, K. Ieki, \emph{et al}., Phys. Lett. B \textbf{346}, 9 (1995).
 \bibitem{Bastin07} B. Bastin, S. Gr\'{e}vy, D. Sohler, \emph{et al}., Phys. Rev. Lett. \textbf{99}, 022503 (2007).
 \bibitem{Grevy04} S. Gr\'{e}vy, J. C. Ang\'{e}lique, P. Baumann, \emph{et al}., Phys. Lett. B \textbf{594}, 252 (2004).
 \bibitem{Sorlin93} O. Sorlin, D. Guillemaud-Mueller, A. C. Mueller, \emph{et al}., Phys. Rev. C \textbf{47}, 2941 (1993).
 \bibitem{Scheit96} H. Scheit, T. Glasmacher, B. A. Brown, \emph{et al}., Phys. Rev. Lett. \textbf{77}, 3967 (1996).
 \bibitem{Glasmacher97} T. Glasmacher, B. A. Brown, M. J. Chromik, \emph{et al}., Phys. Lett. B \textbf{395}, 163 (1997).
 \bibitem{Sohler02} D. Sohler, Zs. Dombr\'{a}di, and J. Tim\'{a}r, \emph{et al}., Phys. Rev. C \textbf{66}, 054302 (2002).
 \bibitem{Gade05} A. Gade, D. Bazin, C. A. Bertulani, \emph{et al}., Phys. Rev. C \textbf{71}, 051301(R) (2005).
 \bibitem{Grevy05} S. Gr\'{e}vy, F. Negoita, I. Stefan, \emph{et al}., Eur. Phys. J. A \textbf{25}, 111 (2005).
 \bibitem{Gaudefroy06} L. Gaudefroy, O. Sorlin, D. Beaumel, \emph{et al}., Phys. Rev. Lett. \textbf{97}, 092501 (2006).
 \bibitem{Force10} C. Force, S. Gr\'{e}vy, L. Gaudefroy, \emph{et al}., Phys. Rev. Lett. \textbf{105}, 102501 (2010).

 \bibitem{Arumugam05} P. Arumugam, B. K. Sharma, and S. K. Patra, Phys. Rev. C \textbf{71}, 064308 (2005).
 \bibitem{Oertzen06} W. von. Oertzen, M. Freer and Y. Kanada-En'yo, Phys. Rep. \textbf{432}, 43-113 (2006).
 \bibitem{Beck11} C. Beck, P. Papka, F. Azaiez, \emph{et al}., Acta Phys. Pol. B \textbf{42}, 747 (2011).
 \bibitem{Ebran14} J. P. Ebran, E. Khan, T. Nik\v{s}i\'{c}, and D. Vretenar,
                   Phys. Rev. C \textbf{80}, 054329 (2014).
 \bibitem{Yao14b} J. M. Yao, N. Itagaki, and J. Meng, Phys. Rev. C \textbf{90}, 054307 (2014). \bibitem{Caurier05} E. Caurier, G. Mart\'{i}nez-Pinedo, F. Nowacki, A. Poves, and A. P. Zuker,
                     Rev. Mod. Phys. \textbf{77}, 427 (2005).
 \bibitem{Otsuka01-3} T. Otsuka, M. Honma, T. Mizusaki, N. Shimizu, and Y. Utsuno,
                      Prog. Part. Nucl. Phys. \textbf{47}, 319 (2001).
 \bibitem{Brown88} B. A. Brown and B. H. Wildenthal,
                   Annu. Rev. Nucl. Part. Sci. \textbf{38}, 29 (1988).
 \bibitem{Wildenthal80} B. H. Wildenthal and W. Chung,
                        Phys. Rev. C \textbf{22}, 2260(R) (1980).
 \bibitem{Kaneko11} K. Kaneko, Y. Sun, T. Mizusaki, and M. Hasegawa,
                    Phys. Rev. C \textbf{83}, 014320 (2011).
 \bibitem{Brown01} B. A. Brown, Prog. Part. Nucl. Phys. \textbf{47}, 517 (2001).
 \bibitem{Utsuno99} Y. Utsuno, T. Otsuka, T. Mizusaki, and M. Honma,
                    Phys. Rev. C \textbf{60}, 054315 (1999).

 \bibitem{Bender03} M. Bender, P.-H. Heenen, and P.-G. Reinhard,
                    Rev. Mod. Phys. \textbf{75}, 121 (2003).
 \bibitem{Rodriguez-Guzman00} R. Rodr\'{i}guez-Guzm\'{a}n, J. L. Egido, and L. M. Robledo, Phys. Rev. C \textbf{62}, 054319 (2000);
                              Phys. Rev. C \textbf{65}, 024304 (2002); Phys. Lett. B \textbf{474}, 15 (2000); Nucl. Phys. A \textbf{709}, 201 (2002).
 \bibitem{Valor00} A. Valor, P.-H. Heenen, and P. Bonche,
                   Nucl. Phys. A \textbf{671}, 145 (2000).
 \bibitem{Niksic06} T. Nik\v{s}i\'{c}, D. Vretenar, and P. Ring,
                    Phys. Rev. C \textbf{73}, 034308 (2006); \textbf{74}, 064309 (2006).
 \bibitem{Bender03PRC} M. Bender, H. Flocard, and P.-H. Heenen, Phys. Rev. C \textbf{68}, 044321 (2003).
 \bibitem{Rodriguez2010} T. R. Rodr\'{\i}guez and J. L. Egido, Phys. Rev. C \textbf{81}, 064323 (2010).
 \bibitem{Bender08} M. Bender, P.-H. Heenen, Phys. Rev. C \textbf{78}, 024309 (2008). \bibitem{MRCDFT} J. M. Yao, J. Meng, P. Ring, and D. Pena Arteaga, Phys. Rev. C \textbf{79}, 044312 (2009);
                  J. M. Yao, J. Meng, P. Ring, and D. Vretenar, Phys. Rev. C \textbf{81}, 044311 (2010).
 \bibitem{Bertsch07} G. F. Bertsch, M. Girod, S. Hilaire, J.-P. Delaroche, H. Goutte, and S. P\'eru,
                     Phys. Rev. lett. \textbf{99}, 032502 (2007).
 \bibitem{Delaroche10} J.-P. Delaroche, M. Girod, J. Libert, H. Goutte, S. Hilaire, S. P\'{e}ru, N. Pillet, and G. F. Bertsch, Phys. Rev. C \textbf{81}, 014303 (2010).
 \bibitem{Lu15} K. Q. Lu, Z. X. Li, Z. P. Li, J. M. Yao, and J. Meng,
                Phys. Rev. C \textbf{91}, 027304 (2015).
 \bibitem{Bender05} M. Bender, G. F. Bertsch, and P.-H. Heenen,
                    Phys. Rev. Lett. \textbf{94}, 102503 (2005).
 \bibitem{Bender06a} M. Bender, P. Bonche, and P.-H. Heenen,
                    Phys. Rev. C \textbf{74}, 024312 (2006).
 \bibitem{Sabbey07} B. Sabbey, M. Bender, G. F. Bertsch, and P. H. Heenen,
                    Phys. Rev. C \textbf{75}, 044305 (2007).
 \bibitem{Rodr14} T. R. Rodr\'{\i}guez, A. Arzhanov, G. Mart\'{\i}nez-Pinedo, Phys. Rev. C \textbf{91}, 044315 (2015).
 \bibitem{Kanada95} Y. Kanada-En'yo and H. Horiuchi, Prog. Theor. Phys. \textbf{93}, 115 (1995).
 \bibitem{Kimura04} M. Kimura, Phys. Rev. C \textbf{69}, 044319 (2004).
 \bibitem{Kanada12} Y. Kanada-En'yo, M. Kimura, and A. Ono, Prog. Theor. Exp. Phys. \textbf{2012}, 01A202 (2012).
 \bibitem{Pillet08} N. Pillet, J.-F. Berger, and E. Caurier, Phys. Rev. C \textbf{78}, 024305 (2008).
 \bibitem{Bloas14} J. Le Bloas, N. Pillet, M. Dupuis, J. M. Daugas, L. M. Robledo, C. Robin, and V. G. Zelevinsky,
                   Phys. Rev. C \textbf{89}, 011306(R) (2014).
 \bibitem{Yao13} J. M. Yao, H. Mei, Z. P. Li,
                 Phys. Lett. B \textbf{459}, 723 (2013).
 \bibitem{Yao14} J. M. Yao, K. Hagino, Z. P. Li, J. Meng, and P. Ring, Phys. Rev. C \textbf{89}, 054306 (2014).
 \bibitem{Yao11-2} J. M. Yao, J. Meng, P. Ring, Z. X. Li, Z. P. Li, and K. Hagino,
                   Phys. Rev. C \textbf{84}, 024306 (2011).
 \bibitem{Yao11} J. M. Yao, H. Mei, H. Chen, J. Meng, P. Ring, and D. Vretenar,
                 Phys. Rev. C \textbf{83}, 014308 (2011).
 \bibitem{Wang14} Y. Wang, J. Li, J. B. Lu, and J. M. Yao,
                  Prog. Theor. Exp. Phys. \textbf{2014}, 113D03 (2014).
 \bibitem{Burvenich02} T. Burvenich, D. G. Madland, J. A. Maruhn, and P. G. Reinhard,
                       Phys. Rev. C \textbf{65}, 044308 (2002).
 \bibitem{Zhao2010} P. W. Zhao, Z. P. Li, J. M. Yao, and J. Meng, Phys. Rev. C \textbf{82}, 054319 (2010).
 \bibitem{Meng2013} J. Meng, J. Peng, S. Q. Zhang, and P. W. Zhao, Front. Phys. \textbf{8}, 55-79 (2013).
 \bibitem{Meng1998} J. Meng, Nucl. Phys. A \textbf{635}, 3-42 (1998).
 \bibitem{Long2004} W. H. Long, J. Meng, N. Van Giai, and S. G. Zhou, Phys. Rev. C \textbf{69}, 034319 (2004).
 \bibitem{Meng2006} J. Meng, H. Toki, S. G. Zhou, S. Q. Zhang, W. H. Long, and L. S. Geng, Prog. Part. Nucl. Phys. \textbf{57}, 470-563 (2006).
 \bibitem{Ring80} P. Ring, P. Schuck, The Nuclear Many-Body Problem, Spinger, Heidelberg, 1980.
 \bibitem{Griffin57} J. J. Griffin and J. A. Wheeler,
                     Phys. Rev. \textbf{108}, 311-327 (1957).
 \bibitem{Hill53} D. L. Hill and J. A. Wheeler, Phys. Rev. \textbf{89}, 1102 (1953).
 \bibitem{Yao10} J. M. Yao, J. Meng, P. Ring, and D. Vretenar, Phys. Rev. C \textbf{81}, 044311 (2010).
 \bibitem{Yao06} J. M. Yao, H. Chen, and J. Meng, Phys. Rev. C \textbf{74}, 024307 (2006).
 \bibitem{Krieger90} S. J. Krieger, P. Bonche, H. Flocard, P. Quentin, M. S. Weiss, Nucl. Phys. A \textbf{517}, 275 (1990).
 \bibitem{Fomenko70} V. N. Fomenko, J. Phys. A: Gen. Phys. \textbf{3}, 8 (1970).
 \bibitem{Robledo09} L. M. Robledo, Phys. Rev. C \textbf{79}, 021302(R) (2009).
 \bibitem{Bender09} M. Bender, T. Duguet, and D. Lacroix, Phys. Rev. C \textbf{79}, 044319 (2009).
 \bibitem{Tomas11} T. R. Rodr\'iguez and J. L. Egido, Phys. Rev. C \textbf{84}, 051307(R) (2011).
 \bibitem{Santiago-Gonzalez11} D. Santiago-Gonzalez, I. Wiedenh\"{o}ver, V. Abramkina,  \emph{et al.},
                               Phys. Rev. C \textbf{83}, 061305(R) (2011).
 \bibitem{Li11}  Z. P. Li, J. M. Yao, D. Vretenar, T. Nik\v{s}i\'{c}, H. Chen, and J. Meng,
                 Phys. Rev. C \textbf{84}, 054304 (2011).
 \bibitem{Utsuno2015} Y. Utsuno, N. Shimizu, T. Otsuka, T. Yoshida, and Y. Tsunoda, Phys. Rev. Lett. \textbf{114}, 032501 (2015).
 \bibitem{Caceres12} L. C\'{a}ceres, D. Sohler, S. Gr\'{e}vy, \emph{et al.}, Phys. Rev. C \textbf{85}, 024311 (2012).
 \bibitem{Gori09a} S. Goriely, N. Chamel, and J. M. Pearson, Phys. Rev. Lett. \textbf{102}, 152503 (2009).
 \bibitem{Zhang14} Q. S. Zhang, Z. M. Niu, Z. P. Li, J. M. Yao, and J. Meng,
                   Front. Phys. \textbf{9}, 529 (2014).
 \bibitem{Gori05} S. Goriely, M. Samyn, J. M. Pearson, and M. Onsi, Nucl. Phys. A \textbf{750}, 425 (2005).
 \bibitem{Cham08} N. Chamel, S. Goriely, and J. M. Pearson, Nucl. Phys. A \textbf{812}, 72 (2008).
 \bibitem{Bend05} M. Bender, G. F. Bertsch, and P.-H. Heenen, Phys. Rev. Lett. \textbf{94}, 102503 (2005).
 \bibitem{Bend06}M. Bender, G. F. Bertsch, and P.-H. Heenen, Phys. Rev. C \textbf{73}, 034322 (2006).
 \bibitem{Gori09b} S. Goriely, S. Hilaire, M. Girod, and S. P\'{e}ru, Phys. Rev. Lett. \textbf{102}, 242501 (2009).
 \bibitem{Audi03} G. Audi, A. H. Wapstra, and C. Thibault, Nucl. Phys. A \textbf{729}, 337 (2003).
 \bibitem{Yao12} J. M. Yao, S. Baroni, M. Bender, and P.-H. Heenen, Phys. Rev. C \textbf{86}, 014310 (2012).
 \bibitem{Yao15} J. M. Yao, M. Bender, and P.-H. Heenen, Phys. Rev. C \textbf{91}, 024301 (2015).
 \bibitem{Richter03} W. A. Richter, B. A. Brown, Phys. Rev. C \textbf{67}, 034317 (2003).
 \bibitem{Yordanov12} D. T. Yordanov, M. L. Bissell, K. Blaum, \emph{et al}.,
                      Phys. Rev. Lett. \textbf{108}, 042504 (2012).
 \bibitem{Angeli13} I. Angeli, K. P. Marinova,
                    Atomic Data and Nuclear Data Tables \textbf{99}, 69-95 (2013).
 \bibitem{Grasso09} M. Grasso, L. Gaudefroy, E. Khan, T. Nik\v{s}i\'{c}, J. Piekarewicz, O. Sorlin, N. Van Giai, and D. Vretenar, Phys. Rev. C \textbf{79}, 034318 (2009).
 \bibitem{Wu14} X. Y. Wu, J. M. Yao, and Z. P. Li, Phys. Rev. C \textbf{89}, 017304 (2014).
 \bibitem{Hara82} K. Hara, A. Hayashi, and P. Ring, Nucl. Phys. A \textbf{385}, 14 (1982).
 \bibitem{Bonche90} P. Bonche, J. Dobaczewski, H. Flocard, P.-H. Heenen, and J. Meyer, Nucl. Phys. A \textbf{510}, 466 (1990).
 \bibitem{NNDC} National Nuclear Data Center, Brookhaven National Laboratory, http://www.nndc.bnl.gov/.
 \bibitem{Yao09} J. M. Yao, J. Meng, P. Ring,  and D. Pena Arteaga, Phys. Rev. C \textbf{79}, 044312 (2009).
 \bibitem{Marinova11} K. Marinova, W. Geithner, M. Kowalska, K. Blaum, S. Kappertz, M. Keim, S. Kloos, G. Kotrotsios,
                      Phys. Rev. C \textbf{84}, 034313 (2011).
 \bibitem{Takeuchi12} S. Takeuchi, M. Matsushita, N. Aoi, \emph{et al}.,
                      Phys. Rev. Lett. \textbf{109}, 182501 (2012).
 \bibitem{Doornenbal13} P. Doornenbal, H. Scheit, S. Takeuchi, \emph{et al}.,
                        Phys. Rev. Lett. \textbf{111}, 212502 (2013).
 \bibitem{Borrajo15} M. Borrajo, T. R. Rodr\'{\i}guez and J. L. Egido,
                     Phys. Lett. B \textbf{746}, 341(2015).
 \bibitem{Pritychenko14} B. Pritychenko, M. Birch, B. Singh, M. Horoi,
                        arXiv:1312.5975v4 [nucl-th] (2014).
 \bibitem{Stone14} N. J. Stone,
                   INDC International Nuclear Data Committee, INDC(NDS)-0658, IAEA (2014).
 \bibitem{CEA} S. Hilaire and M. Girod,
 [http://www-phynu.cea.fr/science\underline{~}en\underline{~}ligne/carte\underline{~}potentiels \underline{~}microscopiques/carte\underline{~}potentiel\underline{~}nucleaire.htm].
 \bibitem{Cottle02} P. D. Cottle, Z. Hu, B. V. Pritychenko, \emph{et al}.,
                    Phys. Rev. Lett. \textbf{88}, 17 (2002).
 \bibitem{ALEXANDER82} T. K. Alexander, G. C. Ball, W. G. Davies, J. S. Forster, I. V. Mitchell, H.-B. Mak,
                       Phys. Lett. B \textbf{113}, 132 (1982).
 \bibitem{Khandaker91} M. A. Khandaker, J. J. Kelly, P. Boberg, \emph{et al}.,
                       Phys. Rev. C \textbf{44}, 1978 (1991).
 \bibitem{Kennedy92} M. A. Kennedy, P. D. Cottle, and K. W. Kemper,
                     Phys. Rev. C \textbf{46}, 1811 (1992).
 \bibitem{Kelley97} J. H. Kelley, T. Suomij\"{a}rvi, S. E. Hirzebruch, \emph{et al}.,
                    Phys. Rev. C \textbf{56}, R1206 (1997).
 \bibitem{Alamanos99} N. Alamanos, A. Pakou, A. Lagoyannis, A. Musumarra,
                      Nucl. Phys. A \textbf{660}, 406 (1999).
 \bibitem{Takeuchi09} S. Takeuchi, N. Aoi, T. Motobayashi, \emph{et al}., Phys. Rev. C \textbf{79}, 054319 (2009).
 \bibitem{Michimasa14} S. Michimasa, Y. Yanagisawa, K. Inafuku, \emph{et al}.,
                       Phys. Rev. C \textbf{89}, 054307 (2014).
 \bibitem{Li74} G. C. Li, M. R. Yearian, and I. Sick,
                Phys. Rev. C \textbf{9}, 1861 (1974).
 \bibitem{Brain77} S. W. Brain, A. Johnston, W. A. Gillespie, E. W. Lees, and R. P. Singhal,
                   J. Phys. G: Nucl. Phys. \textbf{3}, 821 (1977).
 \bibitem{Mei15-2} H. Mei, K. Hagino, J. M. Yao, and T. Motoba, Phys. Rev. C \textbf{91}, 064305 (2015).

\end{thebibliography}
\end{document}